\documentstyle[12pt]{article}
\renewcommand{\theequation}{\arabic{equation}}
\newcommand{\EQ}{\begin{equation}}
\newcommand{\EN}{\end{equation}}

\newcommand{\ket}[1]{\left|#1\right\rangle}      

\newcommand{\n}{\noindent}

\newcommand{\bear}{\begin{eqnarray}}
\newcommand{\ear}{\end{eqnarray}}
\newcommand{\bt} { \begin{tabular} }
\newcommand{\et}{ \end{tabular} }
\newcommand{\bc} { \begin{center} }
\newcommand{\ec}{ \end{center} }

\newcommand{\btb} { \begin{table} }
\newcommand{\etb}{ \end{table} }

\begin{document}

\topmargin 0pt
\oddsidemargin 5mm
\newcommand{\NP}[1]{Nucl.\ Phys.\ {\bf #1}}
\newcommand{\PL}[1]{Phys.\ Lett.\ {\bf #1}}
\newcommand{\NC}[1]{Nuovo Cimento {\bf #1}}
\newcommand{\CMP}[1]{Comm.\ Math.\ Phys.\ {\bf #1}}
\newcommand{\PR}[1]{Phys.\ Rev.\ {\bf #1}}
\newcommand{\PRL}[1]{Phys.\ Rev.\ Lett.\ {\bf #1}}
\newcommand{\MPL}[1]{Mod.\ Phys.\ Lett.\ {\bf #1}}
\newcommand{\JETP}[1]{Sov.\ Phys.\ JETP {\bf #1}}
\newcommand{\TMP}[1]{Teor.\ Mat.\ Fiz.\ {\bf #1}}

\renewcommand{\thefootnote}{\fnsymbol{footnote}}

\newpage
\setcounter{page}{0}
\begin{titlepage}
\begin{flushright}
UFSCARF-TH-04-15
\end{flushright}
\vspace{0.5cm}
\begin{center}
{\large $R$-matrices and Spectrum of Vertex Models based on Superalgebras}\\
\vspace{1cm}
{\large W. Galleas and M.J. Martins } \\
\vspace{1cm}
{\em Universidade Federal de S\~ao Carlos\\
Departamento de F\'{\i}sica \\
C.P. 676, 13565-905~~S\~ao Carlos(SP), Brasil}\\
\end{center}
\vspace{0.5cm}

\begin{abstract}
In this paper we investigate trigonometric vertex models associated with solutions of the Yang-Baxter
equation which are invariant relative to $q$-deformed superalgebras $sl(r|2m)^{(2)}$,
$osp(r|2m)^{(1)}$  and
$osp(r=2n|2m)^{(2)}$.
The associated $R$-matrices are presented in terms of the standard
Weyl basis  making possible the formulation of the quantum
inverse scattering method for these lattice models.  This allowed us to derive  
the eigenvectors and the eigenvalues of the corresponding transfer matrices 
as well as explicit expressions for the Bethe ansatz equations.

\end{abstract}

\vspace{.15cm}
\centerline{PACS numbers:  05.50+q, 02.30.IK}
\vspace{.1cm}
\centerline{Keywords: Algebraic Bethe Ansatz, Lattice Models}
\vspace{.15cm}
\centerline{May 2004}
\end{titlepage}

\renewcommand{\thefootnote}{\arabic{footnote}}

\section{Introduction}

In the course of years it has become clear that classical vertex models of statistical mechanics are ideal
paradigm of the theory of two-dimensional integrable systems \cite{BAX}. It turns out that a $R$-matrix satisfying the
Yang-Baxter relation generates in a natural manner the Boltzmann weights of an exactly solved vertex model.
An important family of such models are given by the trigonometric solutions of the Yang-Baxter equation
connected with the fundamental representation of generic $q$-deformed Lie algebras \cite{BA,JI}
and Lie superalgebras \cite{BA1,ZH}.

The physical understanding of vertex models includes necessarily the exact
diagonalization of their transfer matrices, which can provide us information about
the free energy behaviour
and on the nature of the elementary excitations.
This step has been successfully
achieved for standard Lie algebras either by the analytical Bethe ansatz \cite{RE},
a phenomenological technique yielding us solely the transfer matrix eigenvalues,
or through the quantum inverse scattering
method \cite{DEV,MA} which gives us also the eigenvectors. The latter is
a more powerful mathematical approach, offering us the foundation for studying two-dimensional
vertex models from first principles which culminated in an algebraic formulation of
the Bethe ansatz \cite{FA,KO}. The majority of
the algebraic Bethe ansatz results for
superalgebras, however, have been concentrated on the rational $q \rightarrow 1$ limit of
deformed universal
enveloping algebras associated with the $sl(r|m)$ \cite{KUL,GE} and $osp(r|2m)$ \cite{MAR} symmetries.
A similar algebraic program
for $q$-deformed superalgebras is still very far from being completed though some progress  already appeared
in the literature. Most of it is associated with the $U_q[sl(r|m)^{(1)}]$ superalgebra, whose $R$-matrix \cite{CH}
elements are the statistical weights of the Perk-Schultz vertex model \cite{PS}, and the corresponding
transfer matrix can be diagonalized by a graded version of the
nested algebraic Bethe ansatz developed originally by Kulish \cite{KUL}.
By way of contrast,
the other
superalgebras  have been studied on a rather case by case basis
and representative examples are the vertex models
associated with certain $q$-deformations of the
$osp(1|2)$ \cite{KR} and
$osp(2|2)$  \cite{ITO,BRA,MAR1} symmetries.

A unified algebraic Bethe ansatz solution of the fundamental vertex models invariant relative to
affine $q$-deformed Lie superalgebras is  indeed
a long-standing open problem
in the field of integrable systems. In order to establish this formulation it is
indispensable to have at hand explicit  expressions  for the
$R$-matrices much like that presented by Jimbo for nonexceptional Lie algebras \cite{JI}.
Despite of recent advances \cite{ZHA,DEL,DEL1,MZ} in developing methods for  constructing solutions
of the Yang-Baxter equation, the results for the $R$-matrices are usually written in terms of projection operators
that still need to be calculated in a convenient orthonormal coordinates. This unsatisfactory
situation is probably related to the fact that Lie superalgebras possess a more involved representation theory
as compared with  ungraded algebras \cite{KAC}. This is particularly complicated for twisted superalgebras,
making it difficult  even to carry  on
an intuitive analysis such as the analytical Bethe ansatz \cite{TE}.

The purpose of this paper is to start to bridge this gap, by presenting the quantum inverse scattering formulation
for the $U_q[sl(r|2m)^{(2)}]$, $U_q[osp(r|2m)^{(1)}]$  and
$U_q[osp(r=2n|2m)^{(2)}]$
vertex models.  We recall that
the symbol $\sigma$ in $U_q[{\cal{G}}^{(\sigma)}]$ refers
to the type of automorphism admitted
by the superalgebra $\cal{G}$ and their  explicit  forms have been summarized in the
work pioneered  by Bazhanov and Shadrikov \cite{BA1}.
We have organized this paper as follows.
In next section we present
the $R$-matrices of these lattice models in terms of the elementary Weyl basis,
paving the way for a Bethe ansatz analysis. In section 3 we describe the essential
tools to solve the eigenvalue problem for the corresponding transfer matrices by the algebraic Bethe ansatz
approach. We use this knowledge in section 4 to present explicit expressions
for the eigenvalues and the Bethe ansatz
equations. Our conclusions are discussed in section 5.
In four appendices we summarize  the crossing matrices, extra commutation 
rules and technical details concerning
the nested Bethe ansatz analysis.

\section{The quantum $R$-matrices}

In this section we shall present some solutions of the Yang-Baxter equation
\begin{equation}
\label{yb}
R_{12}(\lambda) R_{13}(\lambda+\mu) R_{23}(\mu)=R_{23}(\mu) R_{13}(\lambda+\mu) R_{12}(\lambda),
\end{equation}
where the $R$-matrix $R_{ab}(\lambda)$ act on the tensor product of two $Z_2$ graded vector spaces $V_a$ and $V_b$.

In general a graded vector space $V$ is defined by the direct sum $V^{(0)} \oplus V^{(1)}$ where $V^{(0)}$
and $V^{(1)}$ represents
its even and odd subspaces. The $\alpha$-th degree of freedom of these subspaces are
distinguished  by their
Grassmann parity which is a function $p_{\alpha}$ with values in the group $Z_2$,
\begin{equation}
p_{\alpha}=\cases{
0 \;\;\; \mbox{for} \;\; \alpha \;\; \mbox{even } \;\; \cr
1 \;\;\; \mbox{for} \;\; \alpha \;\; \mbox{odd } \;\; \cr }
\end{equation}

We recall that the tensor products in Eq.(\ref{yb}) take into account the gradation of the respective
subspaces. This means that the matrix elements of Eq.(\ref{yb}) will depend crucially on the parities of the
coordinates, see refs. \cite{KUL,GE} for detailed discussion.  It is possible, however, to define a new
matrix ${\check{R}}_{ab}(\lambda)$ satisfying a different relation that is insensitive to grading, namely
\begin{equation}
\label{ybr}
{\check{R}}_{12}(\lambda) {\check{R}}_{23}(\lambda+\mu) {\check{R}}_{12}(\mu)=
{\check{R}}_{23}(\mu) {\check{R}}_{12}(\lambda+\mu) {\check{R}}_{23}(\lambda) .
\end{equation}

This  matrix plays
a direct role
in the quantum inverse scattering formulation and it is simply related to $R_{ab}(\lambda)$ by the following
expression
\begin{equation}
\label{checkR}
{\check{R}}_{ab}(\lambda)=P_{ab} R_{ab}(\lambda) ,
\end{equation}
where $P_{ab}= \displaystyle \sum_{\alpha,\beta=1}^{N} (-1)^{p_{\alpha} p_{\beta}} \hat{e}^{(a)}_{\alpha\beta}
\otimes \hat{e}^{(b)}_{\beta\alpha} $ is
the graded permutation operator and the integer $N$ represents the dimension of the
spaces $V_a$. As usual $\hat{e}^{(a)}_{\alpha\beta} \in V_a $ denotes $N \times N$ matrices having only one
non-vanishing element with value 1 at row $\alpha$ and column $\beta$.

In what follows we will exhibit explicit ${\check{R}}$-matrices expressions associated with
$U_q[sl(r|2m)^{(2)}]$, $U_q[osp(r|2m)^{(1)}]$  and
$U_q[osp(r=2n|2m)^{(2)}]$
Lie superalgebras. We remark that the needed informations
about these algebras, including
the forms of the
possible Coxeter automorphisms and the corresponding $R$-matrices in terms of projectors
have been described in ref. \cite{BA1}. Therefore, here
we restrict ourselves in presenting only the main results for the $\check{R}$-matrices in suitable basis
for an algebraic Bethe ansatz analysis.  
In order to obtain explicit formulas
it is convenient to work with a specific grading and we have chosen the
following one
\begin{equation}
\label{grad}
p_{\alpha}^{(l_0)} =\cases{
\displaystyle  1 \;\; \mbox{for} \;\; \alpha=1,\dots ,m \;\;\; \mbox{and} \;\;\; \alpha =r+m+1,\dots ,r+2m \;\; \cr
\displaystyle  0 \;\; \mbox{for} \;\; \alpha=m+1,\dots ,r+m \;\; \cr },
\end{equation}
where we have introduced the 
label $l_0 \equiv (r|2m)$ to emphasize the numbers of even (r) and odd (2m) elements
of the graded vector
space we are considering.

It turns out that the above mentioned quantum $\check{R}$-matrices in the Weyl basis are
\bear
\label{Ryb}
{\check{R}}^{(l_{0})}_{ab}(\lambda) &=&\sum_{\stackrel{\alpha=1}{\alpha \neq \alpha'}}^{N_{0}} a_{\alpha}^{(l_{0})} (\lambda)
\hat{e}^{(a)}_{\alpha \alpha} \otimes \hat{e}^{(b)}_{\alpha \alpha}
+b^{(l_{0})} (\lambda) \sum_{\stackrel{\alpha ,\beta=1}{\alpha \neq \beta,\alpha \neq \beta'}}^{N_{0}}
\hat{e}^{(a)}_{\beta \alpha} \otimes \hat{e}^{(b)}_{\alpha \beta}  \nonumber \\
&& +{\bar{c}}^{(l_{0})} (\lambda) \sum_{\stackrel{\alpha ,\beta=1}{\alpha < \beta,\alpha \neq \beta'}}^{N_{0}} \hat{e}^{(a)}_{\alpha \alpha} \otimes \hat{e}^{(b)}_{\beta \beta}
+c^{(l_{0})} (\lambda) \sum_{\stackrel{\alpha ,\beta=1}{\alpha > \beta,\alpha \neq \beta'}}^{N_{0}} \hat{e}^{(a)}_{\alpha \alpha} \otimes \hat{e}^{(b)}_{\beta \beta} \nonumber \\
&& + \sum_{\alpha ,\beta =1}^{N_{0}} d_{\alpha, \beta}^{(l_{0})} (\lambda)
\hat{e}^{(a)}_{\alpha' \beta} \otimes \hat{e}^{(b)}_{\alpha \beta'}
\ear
where each index $\alpha$ has its conjugated $\alpha^{'}=N_{0}+1-\alpha$ with $N_{0}$ being the dimension of the
graded vector space with $r$ even and $2m$ odd elements.
The Boltzmann weights  $a_{\alpha}^{(l_{0})}(\lambda)$,
$b^{(l_{0})}(\lambda)$, $c^{(l_{0})}(\lambda)$ and ${\bar{c}}^{(l_{0})} (\lambda)$ are determined by
\begin{eqnarray}
\label{bw1}
a_{\alpha}^{(l_{0})} (\lambda) &=&(e^{2 \lambda} -\zeta^{(l_{0})})(e^{2 \lambda (1-p_{\alpha}^{(l_0)})} -q^2 e^{2\lambda p_{\alpha}^{(l_0)}}) \\
b^{(l_{0})} (\lambda) &=&q(e^{2 \lambda} -1)(e^{2 \lambda} -\zeta^{(l_{0})}) \\
c^{(l_{0})} (\lambda) &=&(1-q^2)(e^{2 \lambda} -\zeta^{(l_{0})}) \\
{\bar{c}}^{(l_{0})} (\lambda) &=& e^{2 \lambda} c^{(l_{0})}(\lambda) ,
\end{eqnarray}
while $d_{\alpha\beta}^{(l_{0})}(\lambda)$   has the form
\begin{equation}
d_{\alpha, \beta}^{(l_{0})} (\lambda)=\cases{
\displaystyle  q(e^{2 \lambda} -1)(e^{2 \lambda} -\zeta^{(l_{0})}) +e^{2\lambda}(q^2 -1)(\zeta^{(l_{0})} -1) \;\;\;\;\;\;\;\;\;\;\;\;\;\;\;\; \mbox{for} \;\; \alpha=\beta=\beta' \;\; \cr
\displaystyle  (e^{2 \lambda} -1)\left[ (e^{2 \lambda} -\zeta^{(l_{0})}) (-1)^{p_{\alpha}^{(l_0)}} q^{2 p_{\alpha}^{(l_0)}} +e^{2\lambda}(q^2 -1) \right] \; \;\;\;\;  \;\; \mbox{for} \;\; \alpha=\beta \neq \beta' \;\; \cr
\displaystyle  (q^{2 }-1)\left[ \zeta^{(l_{0})}(e^{2 \lambda} -1)\frac{\epsilon_{\alpha}}{\epsilon_{\beta}} q^{t_{\alpha}-t_{\beta}} -\delta_{\alpha ,\beta'} (e^{2\lambda} -\zeta^{(l_{0})}) \right] \;\;\;\; \; \;\; \mbox{for} \;\; \alpha < \beta \;\;\cr
\displaystyle  (q^{2 }-1) e^{2 \lambda} \left[ (e^{2 \lambda} -1)\frac{\epsilon_{\alpha}}{\epsilon_{\beta}} q^{t_{\alpha}-t_{\beta}} -\delta_{\alpha ,\beta'} (e^{2\lambda} -\zeta^{(l_{0})}) \right] \;\;\; \;\;\;\;\;\; \mbox{for} \;\; \alpha > \beta \;\;\cr } .
\end{equation}

In table 1 we have collected the values of the dimension $N_{0}$ and the dependence of $\zeta^{(l_{0})}$ with the
parameter $q$ for each Lie superalgebra.
The other variables $\epsilon_{\alpha}$ and $t_{\alpha}$ for the
$U_q[osp(2n|2m)^{(2)}]$ are related to the grading by
\begin{eqnarray}
\epsilon_{\alpha} &=& \cases{
\displaystyle (-1)^{-\frac{p_{\alpha}^{(l_0)}}{2}} \;\;\;\;\;\; \;\;\;\;\;\;\; \;\;\;\;\; \mbox{for} \;\; 1 \leq \alpha \leq \frac{N_{0}}{2} \;\; \cr
\displaystyle -(-1)^{\frac{p_{\alpha}^{(l_0)}}{2}} \;\;\;\;\;\;\;\;\;\;\; \;\;\;\;\;\; \mbox{for} \;\; \frac{N_{0}}{2} +1 \leq \alpha \leq N_{0} \;\; \cr } ,
\\
t_{\alpha} &=& \cases{
\alpha - \left[ \frac{1}{2} +p_{\alpha}^{(l_0)} -2\displaystyle{\sum_{\beta=\alpha}^{\frac{N_{0}}{2}} p_{\beta}^{(l_0)}} \right] \;\;\;\;\;\; \;\;\;\;\;\;\;\; \;\; \mbox{for} \;\; 1 \leq \alpha \leq \frac{N_{0}}{2} \;\; \cr
\alpha + \left[ \frac{1}{2} +p_{\alpha}^{(l_0)} -2\displaystyle{\sum_{\beta=\frac{N_{0}}{2}+1}^{\alpha} p_{\beta}^{(l_0)}} \right] \;\;\;\;\;\;\;\;\;\;\; \mbox{for} \;\; \frac{N_{0}}{2}+1 \leq \alpha \leq N_{0} \;\; \cr } ,
\end{eqnarray}

and for the remaining superalgebras we have

\begin{eqnarray}
\epsilon_{\alpha} &=& \cases{
\displaystyle (-1)^{-\frac{p_{\alpha}^{(l_0)}}{2}} \;\; \;\;\;\;\;\;\;\;\; \;\;\;\; \;\;\;\;\;\;\;\;\;\;\; \;\;\;\;\;\;\;\;\;\;\;\;\; \;\; \;\;\mbox{for} \;\; 1 \leq \alpha < \frac{N_{0}+1}{2} \;\; \cr
\displaystyle 1 \;\; \;\;\;\;\;\;\;\;\;\;\;\;\;\;\;\;\;\;\;\;\;\; \;\;\;\; \;\;\;\;\;\;\;\;\;\;\; \;\;\;\;\;\;\;\;\;\;\;\;\; \;\; \;\; \mbox{for} \;\; \alpha=\frac{N_{0}+1}{2} \;\; \cr
\displaystyle (-1)^{\frac{p_{\alpha}^{(l_0)}}{2}} \;\; \;\;\;\;\;\;\; \;\;\;\; \;\;\;\;\;\;\;\;\;\;\;\;\;\;\;\;\;\;\;\;\;\;\; \;\;\;\;\;\;\;\; \mbox{for} \;\; \frac{N_{0}+1}{2} < \alpha \leq N_{0} \;\; \cr },
\\
t_{\alpha} &=& \cases{
\alpha + \left[ \frac{1}{2} -p_{\alpha}^{(l_0)} +2\displaystyle{\sum_{\alpha \leq \beta < \frac{N_{0}+1}{2}} p_{\beta}^{(l_0)}} \right] \;\; \;\;\;\;\;\; \;\; \mbox{for} \;\; 1 \leq \alpha < \frac{N_{0}+1}{2} \;\; \cr
\frac{N_{0}+1}{2} \; \;\;\;\;\;\;\;\;\;\;\;\;\;\;\;\;\;\;\;\;\;\;\;\;\;\;\;\;\;\;\;\;\;\;\;\;\;\;\;\;\;\;\;\;\;\;\;\;\;\;\; \mbox{for} \;\; \alpha = \frac{N_{0}+1}{2} \;\; \cr
\alpha - \left[ \frac{1}{2} -p_{\alpha}^{(l_0)} +2\displaystyle{\sum_{\frac{N_{0}+1}{2} < \beta \leq \alpha} p_{\beta}^{(l_0)}} \right] \;\;\;\;\;\;\;\; \;\; \mbox{for} \;\; \frac{N_{0}+1}{2} < \alpha \leq N_{0} \;\; \cr } .
\label{bwf}
\end{eqnarray}

We note that the $R$-matrix $R_{12}(\lambda )$ defined by Eqs.(\ref{checkR},\ref{Ryb}) satisfies 
important relations besides
the standard properties of regularity and unitarity. One of them is the so-called $PT$ symmetry given by
\begin{equation}
P_{12}R_{12}(\lambda )P_{12}=R^{st_1 st_2}_{12}(\lambda ) ,
\label{PT}
\end{equation}
where the symbol $st_k$ denotes the supertransposition in the space with index $k$. The other  is the
crossing symmetry, namely
\begin{equation}
\label{cross}
R_{12}(\lambda )=\frac{\rho(\lambda)}{\rho(-\lambda-\eta)}
V_{1} R_{12}^{st_2}(-\lambda -\eta )V^{-1}_{1},
\label{CRO}
\end{equation}
where $\rho(\lambda)$ is a convenient normalization, $\eta$ is the crossing parameter and $V$ is an
anti-diagonal matrix. Since the expressions of some of these quantities are sufficiently
cumbersome we have collected them in appendix A.

We would like to point out that our findings for the $\check{R}$-matrices of the 
$U_q[sl(2n+1|2m)^{(2)}]$ 
and $U_q[osp(2n+1|2m)^{(1)}]$  vertex models remains valid even for $n=0$. In this case, however,
such 
$\hat{R}$-matrices  can be related, by transformations that are compatible with the Yang-Baxter
equation, to that of the 
$U_{\tilde{q}}[B_{m}]$ and
$U_{\tilde{q}}[A^{(2)}_{2m}]$ 
vertex models \cite{JI} with $\tilde{q}=-q^{-1}$, respectively. 
\section{The algebraic Bethe ansatz }

Each Yang-Baxter solution presented in last section may be interpreted as the local Boltzmann
weights of an integrable vertex models on
a square lattice of size $L \times L$ \cite{BAX}.  Its corresponding row-to-row transfer matrix
$T^{(l_{0})}(\lambda)$ can be
conveniently written as the supertrace, over an auxiliary space $\cal{A} \equiv \mathrm{C}^{N_{0}}$, of an
operator denominated monodromy matrix ${\cal{T}}^{(l_{0})}(\lambda)$ \cite{KUL,GE}
\EQ
\label{trans}
T^{(l_{0})}(\lambda)= \mathrm{Str}[{\cal{T}}^{(l_{0})}(\lambda)]=\sum_{\alpha =1}^{N_{0}} (-1)^{p_{\alpha}^{(l_0)}}
 {\cal{T}}_{\alpha \alpha}^{(l_{0})}(\lambda)
\EN
where
${\cal{T}}_{\alpha \beta}^{(l_{0})}(\lambda)$ denotes the elements of the monodromy matrix which is
given by the following ordered product of $R$-matrices
\EQ
\label{mono}
{\cal{T}}^{(l_{0})}(\lambda)= R_{{\cal{A}}L}^{(l_{0})}(\lambda)
R_{{\cal{A}}L-1}^{(l_{0})}(\lambda)  \dots
R_{{\cal{A}}1}^{(l_{0})}(\lambda).
\EN

The local weights of the above expression are obtained from Eq.(\ref{Ryb}) by the relation
$R_{{\cal{A}}j}^{(l_{0})}(\lambda)=P_{{\cal{A}}j}
\check{R}_{{\cal{A}}j}^{(l_{0})}(\lambda) $. They are viewed as
$N_{0} \times N_{0}$ matrices on the auxiliary space $\cal{A}$
whose elements are operators acting nontrivially in the $j$-th quantum space
$\displaystyle{\prod_{j=1}^{L} \otimes \mathrm{C}_j^{N_{0}}}$.
The monodromy operator (\ref{mono}) is  a
basic object in the quantum inverse scattering method and with help of the Yang-Baxter equation (\ref{ybr})
one can show that it satisfy the following quadratic algebra
\EQ
\label{yba}
{\check{R}}^{(l_{0})}_{12}(\lambda-\mu) {\cal{T}}^{(l_{0})}(\lambda) \stackrel{s_0}{\otimes}{\cal{T}}^{(l_{0})}(\mu)
={\cal{T}}^{(l_{0})}(\mu) \stackrel{s_0}{\otimes}{\cal{T}}^{(l_{0})}(\lambda) {\check{R}}^{(l_{0})}_{12}(\lambda-\mu) ,
\EN
where the matrix elements of
${\check{R}}_{12}^{(l_{0})}(\lambda-\mu)$ are the weights (\ref{bw1}-\ref{bwf}) defined on the tensor product $\cal{A} \otimes \cal{A}$.
The symbol
$\stackrel{s_0}{\otimes}$ stands for the supertensor product \cite{KUL} with respect to the auxiliary space $\cal{A}$.
We recall that such product between two matrices with elements $A_{ab}$ and $B_{cd}$ should be understood by
$ A
\stackrel{s_0}{\otimes}B = \displaystyle{\sum_{abcd}^{N_{0}}} (-1)^{p_b^{(l_0)}[p_a^{(l_0)}+p_c^{(l_0)}]} A_{ac} B_{bd} \; \hat{e}_{ac} \otimes \hat{e}_{bd}$ .

The Yang-Baxter algebra (\ref{yba}) plays a fundamental role in the solution of the transfer matrix eigenvalue problem,
\EQ
\label{eigenp}
T^{(l_0)}(\lambda) \ket{\Phi} = \Lambda^{(l_0)} (\lambda) \ket{\Phi},
\EN
by means of an exact operator formalism. Other important ingredient is the existence of a pseudovacuum state
$\ket{\Phi_0}$ in which the monodromy matrix acts triangularly. This state help us to identify the off-diagonal
elements of the monodromy matrix as potential creation and annihilation fields. For the vertex models considered
in this paper we can choose $\ket{\Phi_0}$ as the highest weight state vector
\EQ
\ket{\Phi_0} = \prod_{j=1}^{L} \otimes \ket{0}_{j} , ~~
\ket{0}_{j} =
\pmatrix{
1 \cr
0 \cr
\vdots \cr
0 \cr}_{N_{0}} ,
\EN
where
$\ket{0}_{j}$ is the local reference state at the
$j$-th lattice site with $N_{0}$ components. The action of each  operator
$R^{(l_{0})}_{{\cal{A}}j}(\lambda)$ in this state gives
\EQ
\label{triang}
{R}^{(l_{0})}_{{\cal A}j}(\lambda)\ket{0}_{j} =
\pmatrix{
\omega_1^{(l_0)}(\lambda) \ket{0}_j &  \dagger  &  \dagger  &  \dots & \dagger & \dagger  \cr
0  &  \omega_2^{(l_0)}(\lambda) \ket{0}_j &  0  & \dots & 0 & \dagger  \cr
\vdots & \vdots & \ddots & \dots & \vdots & \vdots \cr
0  &  0  &  0 & \dots & \omega_{N_{0}-1}^{(l_0)}(\lambda) \ket{0}_j& \dagger \cr
0  &  0  &  0  & \dots & 0 &  \omega_{N_{0}}^{(l_0)}(\lambda) \ket{0}_j  \cr}_{N_{0} \times N_{0}}
\EN
where the symbol $\dagger$ stands for non-null values and the functions $\omega_{\alpha}^{(l_0)}(\lambda)$ are given by
\begin{equation}
\omega_{\alpha}^{(l_0)}(\lambda) =\cases{
\displaystyle (-1)^{p_1^{(l_0)}} a_{1}^{(l_{0})} (\lambda) \;\;\;\;\;\; \mbox{for} \;\; \alpha=1 \;\; \cr
\displaystyle (-1)^{p_{\alpha}^{(l_0)}} b^{(l_{0})} (\lambda) \;\;\;\;\;\; \mbox{for} \;\; \alpha=2,\dots,N_{0}-1 \;\; \cr
\displaystyle (-1)^{p_{N_0}^{(l_0)}} d_{N_{0},N_{0}}^{(l_{0})} (\lambda) \;\;\; \mbox{for} \;\; \alpha=N_{0} \;\; \cr } .
\end{equation}

To make further progress one needs to seek for an appropriate representation of the monodromy
matrix that is able to distinguish possible creation and annihilation fields. Previous experience
with the vertex models \cite{MAR} whose weights have similar triangular property such as exhibited in Eq.(\ref{triang})
suggests us that a promissing ansatz should be
\EQ
\label{abcdf}
{\cal T}^{(l_0)}(\lambda) =
\pmatrix{
B(\lambda)       &   \vec{B}(\lambda)   &   F(\lambda)   \cr
\vec{C}(\lambda)  &  \hat{A}(\lambda)   &  \vec{B^{*}}(\lambda)   \cr
C(\lambda)  & \vec{C^{*}}(\lambda)  &  D(\lambda)  \cr}_{N_0 \times N_0},
\EN
where $\vec{B}(\lambda)$
($\vec{B^{*}}(\lambda)$) and
$\vec{C^{*}}(\lambda)$
($\vec{C}(\lambda)$) are $(N_0-2)$-component row (column) vectors,
$ \hat{A}(\lambda)  $ is a $ (N_0-2) \times (N_0-2) $ matrix whose elements will be denoted by
$A_{ab}(\lambda)$ and the remaining operators play the role of scalars. Taking into account this representation
and the grading choice (\ref{grad}), the diagonalization of the transfer matrix becomes equivalent to the problem
\begin{equation}
\left[ (-1)^{p_1^{(l_0)}}B(\lambda) +\sum_{a=1}^{N_0-2} (-1)^{p_a^{(l_0)}}\hat{A}_{aa}(\lambda) +(-1)^{p_{N_0}^{(l_0)}}D(\lambda) \right]
\ket{\phi} =\Lambda^{(l_0)} (\lambda) \ket{\phi}.
\end{equation}

Direct comparison between Eq.(\ref{mono}) and Eq.(\ref{triang}) reveals that
$\vec{B}(\lambda)$,
$\vec{B^{*}}(\lambda)$ and $F(\lambda)$ are creation fields with respect to the reference state $\ket{\Phi_0}$.
Furthermore, the diagonal elements of ${\cal{T}}^{(l_{0})}(\lambda)$ satisfy the relations
\EQ
\matrix{ B(\lambda)\ket{\Phi_0} = [\omega_1(\lambda)]^{L}\ket{\Phi_0} & \;\;\;\;\;\; D(\lambda)\ket{\Phi_0} =
[\omega_{N_0}(\lambda)]^{L}\ket{\Phi_0} \cr
A_{aa}(\lambda)\ket{\Phi_0} = [\omega_{a+1}(\lambda)]^{L}\ket{\Phi_0} & \mathrm{for}~~a=1, \dots , N_0-2 \cr },
\EN
as well as the annihilation properties
\EQ
\matrix{ \vec{C}(\lambda)\ket{\Phi_0} = 0 & \; \vec{C^{*}}(\lambda)\ket{\Phi_0} = 0~~
\;\;\;\;\; C(\lambda) \ket{\Phi_0} = 0 \cr
A_{ab}(\lambda)\ket{\Phi_0} = 0 & \mathrm{for}~~ a,b =1, \dots , N_0-2 \;\;\;\; a \neq b \cr },
\EN
implying that the reference state $\ket{\Phi_0}$ is one of the transfer matrix eigenstates whose respective
eigenvalue is
\begin{equation}
\Lambda^{(l_0)}_{0}(\lambda)=(-1)^{p_1^{(l_0)}}[\omega_{1} (\lambda)]^{L} +\displaystyle{\sum_{a=1}^{N_0-2}} 
(-1)^{p_{a+1}^{(l_0)}}[\omega_{a+1} (\lambda)]^{L} +(-1)^{p_{N_0}^{(l_0)}}[\omega_{N} (\lambda)]^{L}.
\end{equation}

Within the algebraic Bethe ansatz approach we now seek for other transfer matrix eigenvectors as linear
combinations of products of creations fields acting on $\ket{\Phi_0}$. In order to do that we need
to find the appropriate set of commutation rules between the diagonal and creation fields which
in principle are encoded in the Yang-Baxter algebra (\ref{yba}).  The
procedure of deriving commutation rules
in a convenient form is similar to that describe in ref.\cite{MAR}, requiring in some cases
the substitution of the exchange rules between the scalar operator $B(\lambda)$
and $F(\mu)$ or the vector field ${\vec{B}}^{*}(\lambda)$  back on the original commutation relations coming
from the algebra (\ref{yba}). A considerable amount of additional work is however necessary to include some
adaptations that take into account the grading structure. For example, the commutation relations
between the diagonal fields and the creation operator ${\vec B}(\lambda)$ are
\begin{eqnarray}
\label{comut1}
B(\lambda) \stackrel{s_1}{\otimes} {\vec B}(\mu)&=&(-1)^{p_{12}^{(l_0)}} \frac{a^{(l_{0})}_{1}(\mu-\lambda)}{b^{(l_{0})}(\mu-\lambda)} {\vec B}(\mu) \stackrel{s_1}{\otimes} B(\lambda)
- (-1)^{p_{12}^{(l_0)}} \frac{{c}^{(l_{0})}(\mu-\lambda)}{b^{(l_{0})}(\mu-\lambda)} {\vec B}(\lambda) \stackrel{s_1}{\otimes} B(\mu) \nonumber \\
\\
D(\lambda) \stackrel{s_1}{\otimes} {\vec B}(\mu)&=&(-1)^{p_{12}^{(l_0)}} \frac{b^{(l_{0})}(\lambda-\mu)}{d^{(l_{0})}_{N,N}(\lambda-\mu)} {\vec B}(\mu) \stackrel{s_1}{\otimes} D(\lambda)
- \frac{d^{(l_{0})}_{N_0,1}(\lambda-\mu)}{d^{(l_{0})}_{N_0,N_0}(\lambda-\mu)} F(\lambda)\stackrel{s_1}{\otimes} {\vec C}^{*}(\mu) \nonumber \\
&+& \frac{{c}^{(l_{0})}(\lambda-\mu)}{d^{(l_{0})}_{N_0,N_0}(\lambda-\mu)} F(\mu)\stackrel{s_1}{\otimes} {\vec C}^{*}(\lambda)
- \frac{
{\vec{\xi}}^{(l_{0})}_{1}(\lambda-\mu)} 
{d^{(l_{0})}_{N_0,N_0}(\lambda-\mu)} 
\cdot 
\left[ {\vec B}^{*}(\lambda) \otimes \hat{A}(\mu) \right]
\nonumber \\
\end{eqnarray}
\begin{eqnarray}
\hat{A}(\lambda) \stackrel{s_1}{\otimes} {\vec B}(\mu) &=& \frac{1}{b^{(l_{0})}(\lambda-\mu)} {\vec B}(\mu) \stackrel{s_1}{\otimes} \hat{A}(\lambda) \cdot {\check{r}}_{12}^{(l_1)}(\lambda-\mu)
- \frac{{\bar{c}}^{(l_{0})}(\lambda-\mu)}{b^{(l_{0})}(\lambda-\mu)} {\vec B}(\lambda) \stackrel{s_1}{\otimes} \hat{A}(\mu) \nonumber \\
&+&\frac{1}{d^{(l_{0})}_{N_0,N_0}(\lambda-\mu)} \left[(-1)^{p_{12}^{(l_0)}} {\vec B}^{*}(\lambda) \stackrel{s_1}{\otimes} B(\mu) + \frac{{\bar{c}}^{(l_{0})}(\lambda-\mu)}{b^{(l_{0})}(\lambda-\mu)} F(\lambda) \stackrel{s_1}{\otimes} {\vec C}(\mu) \right] \otimes
{\vec{\xi}}^{(l_{0})}_{1}(\lambda-\mu) 
\nonumber \\
&+&\frac{1}{b^{(l_{0})}(\lambda-\mu)} \left[ F(\mu) \stackrel{s_1}{\otimes} {\vec C}(\lambda) \right] \otimes
{\vec{\xi}}^{(l_{0})}_{2}(\lambda-\mu)
\end{eqnarray}
where $p_{ab}^{(l_0)}=p_a^{(l_0)}+p_b^{(l_0)}$ and
the symbol $\stackrel{s_1}{\otimes}$ denotes the supertensor product with
new Grassmann parities $p_{\alpha}^{(l_1)}$ related to the previous
ones by $p_{\alpha}^{(l_1)}=p_{\alpha +1}^{(l_0)}, \;\;\; \alpha=1,\dots,N_0-2 $.  Furthermore, the vectors
${\vec{\xi}}^{(l_{0})}_{1}$ and
${\vec{\xi}}^{(l_{0})}_{2}$ are given by
\begin{equation}
{\vec{\xi}}^{(l_{0})}_{1}(\lambda)=\sum_{a=1}^{N_0-2} d^{(l_{0})}_{N_0,a+1} (\lambda) \; \hat{e}_{a} \otimes \hat{e}_{N_0-1-a}
\end{equation}
and
\begin{equation}
{\vec{\xi}}^{(l_{0})}_{2}(\lambda)=\sum_{a=1}^{N_0-2} \left[ d^{(l_{0})}_{1,a+1}(\lambda) - d^{(l_{0})}_{N_0,a+1}(\lambda) \frac{d^{(l_{0})}_{1,N_0}(\lambda)}{d^{(l_{0})}_{N_0,N_0}(\lambda)} \right] \; \hat{e}_{a} \otimes \hat{e}_{N_0-1-a} , \nonumber
\end{equation}
such that $\hat{e}_i$ is a vector of length $N_0-2$ with only one
non-null unitary element at $i$-th position.
The label $l_{\alpha}$ generalizes previous definition, characterizing the
graded vector space with $N_{\alpha}=N_0-2\alpha$ degrees of freedom whose number
of even and odd elements
is determined by the following rule
\begin{equation}
\label{lala}
l_{\alpha} \equiv \cases{ (r|2m-2\alpha)~~ ~~ \;\;\;\;\;\;\;\;\;\;\; \mbox{for} \;\; m \geq \alpha \cr
(r+2m-2\alpha|0)~~ ~~ \;\;\;\; \mbox{for} \;\; 0 \leq m < \alpha \cr },
\end{equation}

The final definition entering the commutation rules is concerned with the auxiliary $\check{R}$-matrix
${\check{r}}_{ab}^{(l_1)}(\lambda)$. It is obtained from Eq.(\ref{Ryb}) by the
expression
\begin{equation}
\label{kkk}
\check{r}_{ab}^{(l_{\alpha})} (\lambda)=
{\kappa}^{(l_{\alpha-1})}(\lambda) \check{R}_{ab}^{(l_{\alpha})}(\lambda) ~~~
{\kappa}^{(l_{\alpha})}(\lambda)=q^{(l_{\alpha})}
\frac{b^{(l_{\alpha})}(\lambda)}{d_{N_{\alpha}, N_{\alpha}}^{(l_{\alpha})}(\lambda)}
\end{equation}
where $q^{(l_{\alpha})}=
(-1)^{p_1^{(l_{\alpha})}} q^{1-2p_1^{(l_{\alpha})} }$ and 
$\check{R}_{ab}^{(l_{\alpha})}(\lambda)$ is the 
$\check{R}$-matrix (\ref{Ryb}) defined in the
graded space labeled by $l_{\alpha}$. 

The other sets of commutation rules necessary in the solutions of the eigenvalue problem are
those between the diagonal fields and the scalar creation operator $F(\lambda)$ as well as among all
the creation fields.  In order to avoid overcrowding this section with extra heavier formulae we have
summarized them in Appendix B. It turns out that
the role analysis of the transfer matrix eigenvalue problem described in ref.\cite{MAR}
can be adjusted to cover the analogous problem (\ref{eigenp})  for the trigonometric
vertex models described in section 2. Considering that this procedure has been well explained
in the above mentioned reference, there is no need
to repeat it here again, and in what follows we will present
only the essential points concerning the properties of the eigenvectors and eigenvalues.
As usual the eigenvectors of the transfer matrix are built up in terms of a linear combination
of products of the many creations operators acting on the pseudo vacuum state $\ket{\Phi_0}$.
They form a
multiparticle state structure
characterized by a set of rapidities $\{ \lambda_j^{(l_1)} \}$, that parameterize the creation fields
and can be written in terms of the following scalar product,
\EQ
\label{vector1}
\ket{\Phi_{m_{l_1}}} =
\vec {\Phi}_{m_{l_1}}(\lambda_{1}^{(l_1)}, \dots ,\lambda_{m_{l_1}}^{(l_1)}) \cdot \vec{\cal{F}} \ket{\Phi_0},
\EN
where the  vector $\vec{\cal{F}} \in \displaystyle{\prod_{j=1}^{m_{l_1}} \otimes \; \mathrm{C}_j^{N_0-2}}$ whose coefficients are
going to be denoted by
${\cal{F}}^{a_{m_{l_1}} \dots a_1}$ and the indices $a_j$ run over $N_0-2$ possible values. The structure of the
vector $\vec {\Phi}_{m_{l_1}}(\lambda_{1}^{(l_1)}, \dots ,\lambda_{m_{l_1}}^{(l_1)})$
obeys the following second order recursion relation
\bear
\label{vector}
\vec {\Phi}_{m_{l_1}}(\lambda_{1}^{(l_1)},\dots,\lambda_{m_{l_1}}^{(l_1)}) & = &
\vec {B}(\lambda_{1}^{(l_1)}) \stackrel{s_1}{\otimes} \vec {\Phi}_{m_{l_1}-1}(\lambda_{2}^{(l_1)},
\dots,\lambda_{m_{l_1}}^{(l_1)}) \nonumber \\
&-&
\sum_{j=2}^{m_{l_1}} (-1)^{p_{12}^{(l_0)}}
\frac{{\vec{\xi}}_{1}^{(l_{0})}(\lambda_{1}^{(l_1)}-\lambda_{j}^{(l_1)})}
{d_{N_0,N_0}^{(l_0)}(\lambda_{1}^{(l_1)}-\lambda_{j}^{(l_1)})}
\prod_{k=2,k \neq j}^{m_{l_1}}
\frac{a_1^{(l_{0})}(\lambda_{k}^{(l_1)}-\lambda_{j}^{(l_1)}) }
{b^{(l_{0})}(\lambda_{k}^{(l_1)}-\lambda_{j}^{(l_1)}) }
\nonumber \\
&+& F(\lambda_1^{(l_1)}) \stackrel{s_1}{\otimes}
\vec {\Phi}_{m_{l_1}-2}(\lambda_{2}^{(l_1)},\dots,\lambda_{j-1}^{(l_1)},\lambda_{j+1}^{(l_1)},\dots,\lambda_{m_{l_1}}^{(l_1)})
\nonumber \\
&\times& B(\lambda_j^{(l_1)})
\prod_{k=2}^{j-1}
\frac{\check{r}^{(l_1)}_{k,k+1}(\lambda_{k}^{(l_1)}-\lambda_{j}^{(l_1)})}
{a_1^{(l_{0})}(\lambda_{k}^{(l_1)}-\lambda_{j}^{(l_1)}) }
\nonumber \\
\ear
where we see that
the vector
${\vec{\xi}}_{1}^{(l_{0})}(\lambda)$
projects out from the linear combination (\ref{vector}) certain
states that describe pair of excitations with the same bare momenta $\lambda_j^{(l_1)}$. It therefore plays the role
of a generalized exclusion rule, forbidding certain states at the same site of lattice.

In order to make the eigenkets defined by Eqs.(\ref{vector1}-\ref{vector})
true eigenvectors of the transfer matrix $T^{(l_{0})}(\lambda)$ it is required that the vector
$\vec{\cal{F}}$ be an eigenstate of a inhomogeneous transfer matrix $\tilde{T}^{(l_1)}(\lambda,\{ \lambda_j^{(l_1)} \})$
whose Boltzmann weights
$r^{(l_1)}_{{\cal{A}}^{(1)}j}(\lambda)$
are directly related to the auxiliary matrix
$\check{r}^{(l_1)}_{ab}(\lambda)$ by
\EQ
r^{(l_1)}_{{\cal{A}}^{(1)}j}(\lambda) =
P_{{\cal{A}}^{(1)}j}\check{r}^{(l_1)}_{{\cal{A}}^{(1)}j}(\lambda)
\EN
where now ${\cal{A}}^{(1)} \in \mathrm{C}^{N_0-2}$, i.e a space with two less
degrees of freedom as compared
with ${\cal{A}}$. As before
$\tilde{T}^{(l_1)}(\lambda,\{ \lambda_j^{(l_1)} \})$ is given in terms of the supertrace of a monodromy matrix over
the space
${\cal{A}}^{(1)}$ by
\EQ
\label{transnest}
\tilde{T}^{(l_1)}(\lambda,\{ \lambda_j^{(l_1)} \}) =Str_{{\cal{A}}^{(1)}} \left[ {r}^{(l_1)}_{{\cal{A}}^{(1)}m_{l_1}}(
\lambda-\lambda_{m_{l_1}}^{(l_1)})
{r}^{(l_1)}_{{\cal{A}}^{(1)}m_{l_1}-1}(\lambda-\lambda_{m_{l_1}-1}^{(l_1)})  \dots
{r}^{(l_1)}_{{\cal{A}}^{(1)}1}(\lambda-\lambda_1^{(l_1)}) \right]
\EN

Following the same kind of arguments explained in ref.\cite{MAR} and considering the form of
our commutations rules we find that
the corresponding eigenvalues  are given by the expression,
\begin{eqnarray}
\label{eigenr}
\Lambda^{(l_{0})} (\lambda) &=& (-1)^{p_1^{(l_0)}}[\omega_{1} (\lambda)]^{L}
\prod_{i=1}^{m_{l_1}} (-1)^{p_{1}^{(l_0)}} \frac{a^{(l_{0})}_{1}(\lambda_{i}^{(l_1)}-\lambda)}{b^{(l_{0})}(\lambda_{i}^{(l_1)}-\lambda)} \nonumber \\
&+& (-1)^{p_{N_0}^{(l_0)}}[\omega_{N_0} (\lambda)]^{L} \prod_{i=1}^{m_{l_1}} (-1)^{p_{1}^{(l_0)}} \frac{b^{(l_{0})}(\lambda - \lambda_{i}^{(l_1)})}{d^{(l_{0})}_{N_0,N_0}(\lambda - \lambda_{i}^{(l_1)})} \nonumber \\
&+&[b^{(l_{0})}(\lambda)]^{L} \tilde{\Lambda}^{(l_1)}(\lambda,
\{\lambda_{j}^{(l_1)}\}) 
\prod_{i=1}^{m_{l_1}} \frac{1}{b^{(l_{0})}(\lambda - \lambda_{i}^{(l_1)})},
\end{eqnarray}
where $\tilde{\Lambda}^{(l_1)}(\lambda,\{\lambda^{(l_1)}_{j}\})$ is the eigenvalue of the
inhomogeneous transfer matrix $\tilde{T}^{(l_1)}(\lambda,\{\lambda^{(l_1)}_{j}\})$, and  provided that the rapidities
$\{\lambda^{(l_1)}_{j}\}$ satisfy the Bethe ansatz equations

\begin{eqnarray}
\label{bar}
&&\left[(-1)^{p_1^{(l_0)}} \frac{a^{(l_{0})}_{1}(\lambda_{i}^{(l_1)})}{b^{(l_{0})}(\lambda_{i}^{(l_1)})} \right]^{L}
a^{(l_{0})}_{1}(0) \nonumber \\
&& \prod_{\stackrel{j=1}{j \neq i}}^{m_{l_1}} (-1)^{p_{1}^{(l_0)}}
b^{(l_{0})}(\lambda_{i}^{(l_1)}-\lambda_{j}^{(l_1)})
\frac{a^{(l_{0})}_{1}(\lambda_{j}^{(l_1)}-\lambda_{i}^{(l_1)})}{b^{(l_{0})}(\lambda_{j}^{(l_1)}-\lambda_{i}^{(l_1)})}
=\tilde{\Lambda}^{(l_1)}(\lambda=\lambda_{i}^{(l_1)},\{\lambda_{j}^{(l_1)}\}) \nonumber \\
&& i=1,\dots,m_{l_1}
\end{eqnarray}

This completes only the first step of the Bethe ansatz analysis
because we still need to determine the eigenvalues
$\tilde{\Lambda}^{(l_1)}(\lambda,\{\lambda^{(l_1)}_{j}\})$.  We have  reached a point which
is typical of nested Bethe
ansatz problems  that are going to be discussed in the next section.

\section{Eigenvalues and Bethe Ansatz Equations}

This section is concerned with the diagonalization of the auxiliary
transfer matrix $\tilde{T}^{(l_1)}(\lambda,\{\lambda^{(l_1)}_{j}\} )$ which will be carried out by
another algebraic Bethe ansatz analysis. The corresponding monodromy matrix can be read of from Eq.(\ref{transnest})
and it is
\begin{equation}
\label{mononest}
{\cal T}^{(l_1)}(\lambda,\{\lambda^{(l_1)}_{j}\})= r^{(l_1)}_{{\mathcal A}^{(1)} m_{l_1}} (\lambda-\lambda^{(l_1)}_{m_{l_1}})  r^{(l_1)}_{{\mathcal A}^{(1)} m_{l_1}-1} (\lambda-\lambda^{(l_1)}_{m_{l_1}-1})
\dots r^{(l_1)}_{{\mathcal A}^{(1)} 1} (\lambda-\lambda^{(l_1)}_{1}),
\end{equation}
that satisfies the following intertwining relation
\EQ
\check{r}^{(l_1)}_{12}(\lambda-\mu)
{\cal T}^{(l_1)}(\lambda,\{\lambda^{(1)}_{j}\})
\stackrel{s_1}{\otimes}
{\cal T}^{(l_1)}(\mu,\{\lambda^{(l_1)}_{j}\})=
{\cal T}^{(l_1)}(\mu,\{\lambda^{(l_1)}_{j}\})
\stackrel{s_1}{\otimes}
{\cal T}^{(l_1)}(\lambda,\{\lambda^{(l_1)}_{j}\})
\check{r}^{(l_1)}_{12}(\lambda-\mu) .
\EN

As long as $ N_1 \geq 3$ the structure of the Boltzmann weights
$r^{(l_1)}_{{\mathcal A}^{(1)} j} (\lambda)$ resembles much that of the original vertex operator
$R^{(l_{0})}_{{\mathcal A} j} $ we have begun with. In this situation, we can proceed by adjusting
the main results of previous section but now with $N_0-2$ degrees of freedom as well as by 
taking into account the presence
of the inhomogeneities $\{ \lambda^{(l_1)}_j \}$. The new
pseudovacuum state $\ket{\Phi_{0}^{(1)}}$ in
which the monodromy matrix (\ref{mononest}) acts triangularly is given by
the following ferromagnetic
state
\begin{equation}
\ket{\Phi^{(1)}_0} = \prod_{j=1}^{m_{l_1}} \otimes \ket{0^{(1)}}_{j} , ~~
\ket{0^{(1)}}_{j} =
\pmatrix{
1 \cr
0 \cr
\vdots \cr
0 \cr}_{N_1} \; ,
\end{equation}
and the action of the vertex operator
$r^{(l_1)}_{{\mathcal A}^{(1)} j}(\lambda)$ on it satisfies the relation
\begin{eqnarray}
\label{tri1}
&& r^{(1)}_{{\cal A}^{(1)} j}(\lambda)\ket{0^{(1)}}_{j}    \nonumber \\
&& =\pmatrix{
\omega^{(l_1)}_1(\lambda)\ket{0^{(1)}}_j  &  \dagger  &  \dagger  &  \dots & \dagger & \dagger  \cr
0  &  \omega^{(l_1)}_2(\lambda) \ket{0^{(1)}}_j &  0  & \dots & 0 & \dagger  \cr
\vdots & \vdots & \ddots & \dots & \vdots & \vdots \cr
0  &  0  &  0 & \dots & \omega^{(l_1)}_{N_1-1}(\lambda) \ket{0^{(1)}}_j & \dagger \cr
0  &  0  &  0  & \dots & 0 &  \omega^{(l_1)}_{N_1}(\lambda) \ket{0^{(1)}}_j \cr}_{N_1 \times N_1} \nonumber \\
\end{eqnarray}
where
the non-null values $\omega^{(l_1)}_{\alpha}(\lambda)$ are now given by
\begin{equation}
\omega^{(l_1)}_{\alpha}(\lambda) =\cases{
\displaystyle \kappa^{(l_{0})}(\lambda) \; (-1)^{p_1^{(l_1)}} a_{1}^{(l_1)} (\lambda) 
\;\;\;\;\;\;\;\; \mbox{for} \;\; \alpha=1 \;\; \cr
\displaystyle \kappa^{(l_{0})}(\lambda) \; (-1)^{p_{\alpha}^{(l_1)}} b^{(l_1)} (\lambda)
\;\;\;\;\;\;\;\; \mbox{for} \;\; \alpha=2,\dots,N_1-1 \;\; \cr
\displaystyle \kappa^{(l_{0})}(\lambda) \; (-1)^{p^{(l_1)}_{N_1}} d_{N_1,N_1}^{(l_1)} (\lambda) 
\;\;\;\;\; \mbox{for} \;\; \alpha=N_1 \;\; \cr }.
\end{equation}

As before, the triangularity property  of the  operator
$r^{(l_1)}_{{\mathcal A}^{(1)} j}(\lambda)$ suggest us to take the following representation for
the corresponding monodromy matrix
\begin{equation}
\label{mono1}
{\cal T}^{(l_1)}(\lambda,\{ \lambda^{(l_1)}_{j} \}) =
\pmatrix{
B^{(1)}(\lambda,\{ \lambda^{(l_1)}_{j} \})       &   \vec{B}^{(1)}(\lambda,\{ \lambda^{(l_1)}_{j} \})   &   F^{(1)}(\lambda,\{ \lambda^{(l_1)}_{j} \})   \cr
\vec{C}^{(1)}(\lambda,\{ \lambda^{(l_1)}_{j} \})  &  \hat{A}^{(1)}(\lambda,\{ \lambda^{(l_1)}_{j} \})   &  \vec{B^{*}}^{(1)}(\lambda,\{ \lambda^{(l_1)}_{j} \})   \cr
C^{(1)}(\lambda,\{ \lambda^{(l_1)}_{j} \})  & \vec{C^{*}}^{(1)}(\lambda,\{ \lambda^{(l_1)}_{j} \})  &  D^{(1)}(\lambda,\{ \lambda^{(l_1)}_{j} \})  \cr}_{N_1 \times N_1} .
\end{equation}

As a consequence of this assumption and property (\ref{tri1}) 
we
find that the elements of this monodromy matrix satisfy the following relations
\begin{eqnarray}
B^{(1)} (\lambda,\{ \lambda^{(l_1)}_{j} \}) \ket{\Phi^{(1)}_0} &=& \prod_{i=1}^{m_{l_1}} \omega^{(l_1)}_{1} (\lambda - \lambda^{(l_1)}_{i} ) \ket{\Phi^{(1)}_0}, \nonumber \\
D^{(1)} (\lambda,\{ \lambda^{(l_1)}_{j} \}) \ket{\Phi^{(1)}_0} &=& \prod_{i=1}^{m_{l_1}} \omega^{(l_1)}_{N_1} (\lambda - \lambda^{(l_1)}_{i} ) \ket{\Phi^{(1)}_0}, \nonumber \\
A^{(1)}_{ab} (\lambda,\{ \lambda^{(l_1)}_{j} \}) \ket{\Phi^{(1)}_0} &=& \delta_{ab} \prod_{i=1}^{m_{l_1}} \omega^{(l_1)}_{a+1} (\lambda - \lambda^{(l_1)}_{i} ) \ket{\Phi^{(1)}_0} \\
&& \; a,b=1,\dots,N_1-2 ,  \nonumber
\end{eqnarray}

\n as well as the annihilation  properties
\begin{eqnarray}
\vec{C}^{(1)} (\lambda,\{ \lambda^{(l_1)}_{j} \}) \ket{\Phi^{(1)}_0} &=& 0 , \nonumber \\
\vec{C^{*}}^{(1)} (\lambda,\{ \lambda^{(l_1)}_{j} \}) \ket{\Phi^{(1)}_0} &=& 0 , \\
C^{(1)} (\lambda,\{ \lambda^{(l_1)}_{j} \}) \ket{\Phi^{(1)}_0} &=& 0 . \nonumber
\end{eqnarray}

To implement the diagonalization of the transfer matrix 
$\tilde{T}^{(l_1)}(\lambda,\{ \lambda^{(l_1)}_{j} \})$ we need
to introduce a second Bethe ansatz whose multiparticle
eigenstates
are going to be parameterized  by a new set of inhomogeneities
$\{ \lambda^{(l_2)}_1, \dots, \lambda^{(l_2)}_{m_{l_2}} \}$.
Clearly, the structure of the commutation relations for the  
elements of the monodromy matrix (\ref{mono1})  as well as the eigenvalue construction 
is similar to that presented in section 3 and appendix B. We basically have to change the 
original Boltzmann weights 
by the corresponding ones related to the $\check{R}$-matrix 
${\check{r}}^{(l_{1})}_{{\mathcal A}^{(1)} j}(\lambda)$,
each operator $\hat{O}(\lambda)$
by its corresponding ${\hat{O}}^{(1)}(\lambda, \{\lambda_j^{(l_1)}) \}$, to replace the parameters 
$\{ \lambda_j^{(l_1)} \}$ by 
$\{ \lambda_j^{(l_2)} \}$  and to substitute the parities $p_{\alpha}^{(l_1)}$ by
$p_{\alpha}^{(l_{2})}= p_{\alpha+1}^{(l_{1})}$ for $\alpha=1, \cdots, N_{1}-2$ in the tensor products. As a consequence of that,
the role 
analysis of the previous section 
can be repeated 
for each step $l_{\alpha}$ whose corresponding  Boltzmann weights
$r^{(l_{\alpha})}_{{\mathcal A}^{(\alpha)} j}(\lambda)$  share the essential
features presented by the original $R$-matrix we started with.
To make our notation clear we stress that we are using
$\tilde{T}^{(l_{\alpha})} (\lambda)$ for the transfer matrix associated
to $r_{ab}^{(l_{\alpha})}(\lambda)$ with eigenvalues $\tilde{\Lambda}^{(l_{\alpha})} (\lambda)$, and
$T^{(l_{\alpha})} (\lambda)$ for the one associated with $R_{ab}^{(l_{\alpha})}(\lambda)$ with eigenvalues
$\Lambda^{(l_{\alpha})} (\lambda)$. Note that these $R$-matrices differs 
only by a multiplicative factor as explained  in Eq.(\ref{kkk}).
In particular, the eigenvalues
at nearest neighbor steps $l_{\alpha}$ and $l_{\alpha+1}$  are going to satisfy
a recurrence relation  similar to that exhibited by
expression (\ref{eigenr}). By taking into account our results 
so far it is not difficult to derive  that such
relation is given by
\begin{eqnarray}
\label{eirec}
&&\Lambda^{(l_{\alpha})}(\lambda , \{ \lambda_{1}^{(l_{\alpha})},\dots, \lambda_{m_{l_{\alpha}}}^{(l_{\alpha})} \})=
(-1)^{p^{(l_{\alpha})}_{1}} \prod_{i=1}^{m_{l_{\alpha}}} (-1)^{p^{(l_{\alpha})}_{1}} a_{1}^{(l_{\alpha})} (\lambda - \lambda_{i}^{(l_{\alpha})})
\prod_{i=1}^{m_{l_{\alpha +1}}} (-1)^{p^{(l_{\alpha})}_{1}} \frac{a_{1}^{(l_{\alpha})} (\lambda_{i}^{(l_{\alpha +1})} -\lambda)}{b^{(l_{\alpha})} (\lambda_{i}^{(l_{\alpha +1})} -\lambda)} \nonumber \\
&&+(-1)^{p^{(l_{\alpha})}_{N_{\alpha}}}  \prod_{i=1}^{m_{l_{\alpha}}} (-1)^{p^{(l_{\alpha})}_{N_{\alpha}}} d_{N_{\alpha},N_{\alpha}}^{(l_{\alpha})} (\lambda - \lambda_{i}^{(l_{\alpha})})
\prod_{i=1}^{m_{l_{\alpha +1}}} (-1)^{p^{(l_{\alpha})}_{N_{\alpha}}} \frac{b^{(l_{\alpha})} (\lambda - \lambda_{i}^{(l_{\alpha +1})})}{d_{N_{\alpha},N_{\alpha}}^{(l_{\alpha})} (\lambda - \lambda_{i}^{(l_{\alpha +1})})} \nonumber \\
&&+\prod_{i=1}^{m_{l_{\alpha}}} b^{(l_{\alpha})} (\lambda - \lambda_{i}^{(l_{\alpha})})
\prod_{i=1}^{m_{l_{\alpha +1}}}  \frac{q^{(l_{\alpha})}}{d_{N_{\alpha},N_{\alpha}}^{(l_{\alpha})} (\lambda - \lambda_{i}^{(l_{\alpha +1})})}
\Lambda^{(l_{\alpha +1})}(\lambda , \{ \lambda_{1}^{(l_{\alpha +1})},\dots, \lambda_{m_{l_{\alpha +1}}}^{(l_{\alpha +1})} \}) , \nonumber \\
\end{eqnarray}
where $p_{\beta}^{(l_{\alpha+1})}= p_{\beta+1}^{(l_{\alpha})}$ for $\beta=1, \dots, N_{\alpha}-2$.

In order to be consistent, we need to set
$\lambda_j^{(l_0)} =0$ for $j=1, \dots, m_{l_0}$ and also to make the identification
$m_{l_0} \equiv L$.  By the same token, the variables 
$\{ \lambda_j^{(l_{\alpha+1})} \}$ that parameterize  the eigenvectors  of the
inhomogeneous transfer matrix $T^{(l_{\alpha})}(\lambda,
\{ \lambda_{1}^{(l_{\alpha})},\dots, \lambda_{m_{l_{\alpha}}}^{(l_{\alpha})} \}) $  are required to satisfy  the
following Bethe ansatz equation
\begin{eqnarray}
\label{beterec}
\prod_{i=1}^{m_{l_{\alpha -1}}} (-1)^{p^{(l_{\alpha})}_{1}} \frac{a_{1}^{(l_{\alpha})}(\lambda_{j}^{(l_{\alpha})} - \lambda_{i}^{(l_{\alpha -1})})}{b^{(l_{\alpha})}(\lambda_{j}^{(l_{\alpha})} - \lambda_{i}^{(l_{\alpha -1})})}=&&
\prod_{i \neq j}^{m_{l_{\alpha }}} q^{(l_{\alpha})} (-1)^{p^{(l_{\alpha})}_{12} } \frac{a_{1}^{(l_{\alpha +1})}(\lambda_{j}^{(l_{\alpha})} - \lambda_{i}^{(l_{\alpha })})}{d_{N_{\alpha},N_{\alpha}}^{(l_{\alpha})}(\lambda_{j}^{(l_{\alpha})} - \lambda_{i}^{(l_{\alpha})})}
\frac{b^{(l_{\alpha })}(\lambda_{i}^{(l_{\alpha})} - \lambda_{j}^{(l_{\alpha })})}{a_{1}^{(l_{\alpha})}(\lambda_{i}^{(l_{\alpha})} - \lambda_{j}^{(l_{\alpha})})} \nonumber \\
&& \times \prod_{i=1}^{m_{l_{\alpha +1}}} (-1)^{p^{(l_{\alpha +1})}_{1}} \frac{a_{1}^{(l_{\alpha +1})}(\lambda_{i}^{(l_{\alpha +1})} - \lambda_{j}^{(l_{\alpha +1})})}{b^{(l_{\alpha +1})}(\lambda_{i}^{(l_{\alpha +1})} - \lambda_{j}^{(l_{\alpha +1})})} .
\end{eqnarray}

These expressions can be iterate, beginning on $\alpha=0$, 
until we reach  the final step $l_f$ and therefore up to
$\alpha=f-1$. The number  of steps necessary 
in such  nested Bethe ansatz construction as well as the underlying 
$R$-matrix $r^{(l_f)}_{{\mathcal A}^{(f)} j}(\lambda)$ will depend much on the family of
the vertex model we are diagonalizing.  In Table 2  we present the index $l_f$
and describe the type of vertex models appearing
in the
lowest Bethe ansatz analysis  for each superalgebra
$sl(r|2m)^{(2)}$,
$osp(r|2m)^{(1)}$  and
$osp(2n|2m)^{(2)}$. For sake of completeness in appendix C we show the explicit expressions
for the matrices
$R^{(l_f)}_{{\mathcal A}^{(f)} j}(\lambda)$ and the corresponding eigenvalues of the
inhomogeneous transfer matrix
$T^{(l_f)}(\lambda,\{\lambda_1^{(l_f)}, \dots, \lambda_{m_{l_{f}}}^{(l_f)} \})$.
We have now reached a point which  all the results can be gathered together
in order to find the eigenvalues expression and the Bethe ansatz equations for the vertex models
presented in section 3.  For instance, we start  with the eigenvalue formula (\ref{eigenr})
and use the recurrence relation (\ref{eirec}) until we reach the problem of
diagonalizing the transfer matrix
$T^{(l_f)}(\lambda,\{\lambda_1^{(l_f)}, \dots, \lambda_{m_{l_{f}}}^{(l_f)} \})$. We then
take into account the results for the eigenvalues of these
vertex systems, which have been collected in appendix C. By following this recipe,
it is straightforward to find the final expressions 
for the eigenvalues of the original vertex models
which we shall begin to list bellow. In order to do that it is
convenient to define the function 
$\displaystyle Q_{\alpha}(\lambda)=\prod_{i=1}^{m_{l_{\alpha}}} \sinh{(\lambda -\lambda_{i}^{(l_{\alpha})})}$
and recalling that we have set $q=e^{i\gamma}$.  The  final results are:

\begin{itemize}
\item $U_{q}[sl(2n+1|2m)^{(2)}]$:
\end{itemize}
\begin{eqnarray}
\label{grupo1}
\Lambda^{(l_{0})} (\lambda) &=& (-1)^{p^{(l_{0})}_{1}} \left[ (-1)^{p^{(l_{0})}_{1}} a_{1}^{(l_{0})}(\lambda) \right]^{L}
\left[ \frac{Q_{1}\left( \lambda - i\frac{\gamma}{2} \right)}{Q_{1}\left( \lambda + i\frac{\gamma}{2} \right)}\right]^{2 p^{(l_{0})}_{1} -1} \nonumber \\
&+& (-1)^{p^{(l_{0})}_{N_{0}}} \left[ (-1)^{p^{(l_{0})}_{N_{0}}} d_{N_{0},N_{0}}^{(l_{0})}(\lambda) \right]^{L}
\left[ \frac{Q_{1}\left( \lambda + i\left(m-n \right)\gamma  +i\frac{\pi}{2} \right)}{Q_{1}\left( \lambda + i\left(m-n-1 \right)\gamma +i\frac{\pi}{2} \right)}\right]^{2 p^{(l_{0})}_{1} -1} \nonumber \\
&+& \left[ b^{(l_{0})} (\lambda)  \right]^{L} \sum_{\alpha=1}^{N_{0} -2} G_{\alpha}(\lambda | \{ \lambda_{j}^{(l_{\beta})} \}) \nonumber \\
\end{eqnarray}
\begin{eqnarray}
&& G_{\alpha}(\lambda | \{ \lambda_{j}^{(l_{\beta})} \}) \nonumber \\
&& =\cases{
\frac{Q_{\alpha} \left(\lambda +i\left(\frac{\alpha +2}{2} \right)\gamma  \right)}{Q_{\alpha} \left(\lambda +i\frac{\alpha}{2} \gamma  \right)}
\frac{Q_{\alpha+1} \left(\lambda +i\left(\frac{\alpha -1}{2} \right)\gamma  \right)}{Q_{\alpha+1} \left(\lambda +i\left(\frac{\alpha +1}{2} \right)\gamma  \right)}
\;\;\;\;\;\;\;\;\;\;\;\;\;\;\;\;\;\;\;\;\;\;\;\;\;\;\;\;\;\;\; \mbox{for} \;\; \alpha = 1,\dots ,m-1 \cr
\frac{Q_{\alpha} \left(\lambda +i\left(\frac{\alpha -2}{2} \right)\gamma  \right)}{Q_{\alpha} \left(\lambda +i\frac{\alpha}{2} \gamma  \right)}
\frac{Q_{\alpha+1} \left(\lambda +i\left(\frac{\alpha +1}{2} \right)\gamma  \right)}{Q_{\alpha+1} \left(\lambda +i\left(\frac{\alpha -1}{2} \right)\gamma  \right)}
\;\;\;\;\;\;\;\;\;\;\;\;\;\;\;\;\;\;\;\;\;\;\;\;\;\;\;\;\;\;\; \mbox{for} \;\; \alpha = m  \cr
\frac{Q_{\alpha} \left(\lambda +i\left(m-1 -\frac{\alpha}{2} \right)\gamma  \right)}{Q_{\alpha} \left(\lambda +i\left( m-\frac{\alpha}{2} \right) \gamma  \right)}
\frac{Q_{\alpha+1} \left(\lambda +i\left(m + \frac{1}{2} -\frac{\alpha}{2} \right)\gamma  \right)}{Q_{\alpha+1} \left(\lambda +i\left(m-\frac{1}{2} - \frac{\alpha}{2} \right)\gamma  \right)}
\;\;\;\;\;\;\;\;\;\;\;\;\;\;\;\;\;\;\;\;\; \mbox{for} \;\; \alpha = m+1,\dots ,m+n-1 \neq m+n \cr
\frac{Q_{\alpha} \left(\lambda +i\left(\frac{m-n -2}{2} \right)\gamma  \right)}{Q_{\alpha} \left(\lambda +i\left( \frac{m-n}{2} \right) \gamma \right)}
\frac{Q_{\alpha} \left(\lambda +i\left(\frac{m-n +1}{2} \right)\gamma +i\frac{\pi}{2} \right)}{Q_{\alpha} \left(\lambda+i \left(\frac{m-n-1}{2} \right)\gamma +i\frac{\pi}{2} \right)}
\;\;\;\;\;\;\;\;\;\;\;\;\;\;\;\;\;\;\;\;\;\; \mbox{for} \;\; \alpha = m+n \cr
G_{\alpha - (m+n)}(-i\frac{\pi}{2} - i(m-n-\frac{1}{2})\gamma -\lambda |- \{ \lambda_{j}^{(l_{\beta})} \}) 
\;\;\;\; \mbox{for} \;\; \alpha = m+n+1,\dots ,2m+2n-1 \cr } \nonumber
\end{eqnarray}

\vskip 1.0cm
\begin{itemize}
\item $U_{q}[sl(2n|2m)^{(2)}]$:
\end{itemize}
\begin{eqnarray}
\Lambda^{(l_{0})} (\lambda) &=& (-1)^{p^{(l_{0})}_{1}} \left[ (-1)^{p^{(l_{0})}_{1}} a_{1}^{(l_{0})}(\lambda) \right]^{L}
\left[ \frac{Q_{1}\left( \lambda - i\frac{\gamma}{2} \right)}{Q_{1}\left( \lambda + i\frac{\gamma}{2} \right)}\right]^{2 p^{(l_{0})}_{1} -1} \nonumber \\
&+& (-1)^{p^{(l_{0})}_{N_{0}}} \left[ (-1)^{p^{(l_{0})}_{N_{0}}} d_{N_{0},N_{0}}^{(l_{0})}(\lambda) \right]^{L}
\left[ \frac{Q_{1}\left( \lambda + i\left(m-n+\frac{1}{2} \right)\gamma +i\frac{\pi}{2} \right)}{Q_{1}\left( \lambda + i\left(m-n-\frac{1}{2} \right)\gamma +i\frac{\pi}{2}  \right)}\right]^{2 p^{(l_{0})}_{1} -1} \nonumber \\
&+& \left[ b^{(l_{0})} (\lambda)  \right]^{L} \sum_{\alpha=1}^{N_{0} -2} G_{\alpha}(\lambda | \{ \lambda_{j}^{(l_{\beta})} \}) \nonumber \\
\end{eqnarray}
\begin{eqnarray}
&& G_{\alpha}(\lambda | \{ \lambda_{j}^{(l_{\beta})} \}) \nonumber \\
&& =\cases{
\frac{Q_{\alpha} \left(\lambda +i\left(\frac{\alpha +2}{2} \right)\gamma  \right)}{Q_{\alpha} \left(\lambda +i\frac{\alpha}{2} \gamma  \right)}
\frac{Q_{\alpha+1} \left(\lambda +i\left(\frac{\alpha -1}{2} \right)\gamma  \right)}{Q_{\alpha+1} \left(\lambda +i\left(\frac{\alpha +1}{2} \right)\gamma  \right)}
\;\;\;\;\;\;\;\;\;\;\;\;\;\;\;\;\;\;\;\;\;\;\;\;\; \mbox{for} \;\; \alpha = 1,\dots ,m-1 \cr
\frac{Q_{\alpha} \left(\lambda +i\left(\frac{\alpha -2}{2} \right)\gamma  \right)}{Q_{\alpha} \left(\lambda +i\frac{\alpha}{2} \gamma  \right)}
\frac{Q_{\alpha+1} \left(\lambda +i\left(\frac{\alpha +1}{2} \right)\gamma  \right)}{Q_{\alpha+1} \left(\lambda +i\left(\frac{\alpha -1}{2} \right)\gamma  \right)}
\;\;\;\;\;\;\;\;\;\;\;\;\;\;\;\;\;\;\;\;\;\;\;\;\; \mbox{for} \;\; \alpha = m \neq m+n-1 \cr
\frac{Q_{\alpha} \left(\lambda +i\left(m-1 - \frac{\alpha}{2} \right)\gamma  \right)}{Q_{\alpha} \left(\lambda +i\left( m-\frac{\alpha}{2} \right) \gamma  \right)}
\frac{Q_{\alpha+1} \left(\lambda +i\left(m + \frac{1}{2} -\frac{\alpha}{2} \right)\gamma  \right)}{Q_{\alpha+1} \left(\lambda +i\left(m-\frac{1}{2} - \frac{\alpha}{2} \right)\gamma  \right)}
\;\;\;\;\;\;\;\;\;\;\;\;\;\;\; \mbox{for} \;\; \alpha = m+1,\dots ,m+n-2 \neq m+n-1 \cr
\frac{Q_{\alpha} \left(\lambda +i\left(\frac{\alpha -2}{2} \right)\gamma  \right)}{Q_{\alpha} \left(\lambda +i\frac{\alpha}{2} \gamma  \right)}
\frac{Q_{\alpha+1} \left(\lambda +i\left(\frac{\alpha +1}{2} \right)\gamma  \right)}{Q_{\alpha+1} \left(\lambda +i\left(\frac{\alpha -1}{2} \right)\gamma  \right)}
\frac{Q_{\alpha+1} \left(\lambda +i\left(\frac{\alpha +1}{2} \right)\gamma  +i\frac{\pi}{2} \right)}{Q_{\alpha+1} \left(\lambda +i\left(\frac{\alpha -1}{2} \right)\gamma +i\frac{\pi}{2} \right)}
\;\;\;\;\;\;\;\;\;\;\;\;\;\;\;\;\;\;\;\;\;\;\;\;\;\; \mbox{for} \;\; \alpha = m +n-1 =m \cr
\frac{Q_{\alpha} \left(\lambda +i\left(m-1 -\frac{\alpha}{2} \right)\gamma  \right)}{Q_{\alpha} \left(\lambda +i\left( m-\frac{\alpha}{2} \right) \gamma  \right)}
\frac{Q_{\alpha+1} \left(\lambda +i\left(m + \frac{1}{2} -\frac{\alpha}{2} \right)\gamma  \right)}{Q_{\alpha+1} \left(\lambda +i\left(m - \frac{1}{2} - \frac{\alpha}{2} \right)\gamma  \right)}
\frac{Q_{\alpha+1} \left(\lambda +i\left(m + \frac{1}{2} -\frac{\alpha}{2} \right)\gamma +i\frac{\pi}{2} \right)}{Q_{\alpha+1} \left(\lambda +i\left(m - \frac{1}{2} - \frac{\alpha}{2} \right)\gamma +i\frac{\pi}{2} \right)}
\;\;\;\;\;\;\;\;\;\;\; \mbox{for} \;\; \alpha = m+n-1 \neq m \cr
G_{\alpha - (m+n-1)}(-i\frac{\pi}{2} -i(m-n)\gamma -\lambda |- \{ \lambda_{j}^{(l_{\beta})} \}) \;\;\;\;\;\;\;\;\;\;\;\;\;\;\;\;\; \mbox{for} \;\; \alpha = m+n,\dots ,2m+2n-2 \cr } \nonumber
\end{eqnarray}

\vskip 1.0cm
\begin{itemize}
\item $U_{q}[osp(2n|2m)^{(1)}]$:
\end{itemize}
\begin{eqnarray}
\Lambda^{(l_{0})} (\lambda) &=& (-1)^{p^{(l_{0})}_{1}} \left[ (-1)^{p^{(l_{0})}_{1}} a_{1}^{(l_{0})}(\lambda) \right]^{L}
\left[ \frac{Q_{1}\left( \lambda - i\frac{\gamma}{2} \right)}{Q_{1}\left( \lambda + i\frac{\gamma}{2} \right)}\right]^{2 p^{(l_{0})}_{1} -1} \nonumber \\
&+& (-1)^{p^{(l_{0})}_{N_{0}}} \left[ (-1)^{p^{(l_{0})}_{N_{0}}} d_{N_{0},N_{0}}^{(l_{0})}(\lambda) \right]^{L}
\left[ \frac{Q_{1}\left( \lambda + i\left(m-n+\frac{3}{2} \right)\gamma \right)}{Q_{1}\left( \lambda + i\left(m-n+\frac{1}{2} \right)\gamma  \right)}\right]^{2 p^{(l_{0})}_{1} -1} \nonumber \\
&+& \left[ b^{(l_{0})} (\lambda)  \right]^{L} \sum_{\alpha=1}^{N_{0} -2} G_{\alpha}(\lambda | \{ \lambda_{j}^{(l_{\beta})} \}) \nonumber \\
\end{eqnarray}
\begin{eqnarray}
&& G_{\alpha}(\lambda | \{ \lambda_{j}^{(l_{\beta})} \}) \nonumber \\
&& =\cases{
\frac{Q_{\alpha} \left(\lambda +i\left(\frac{\alpha +2}{2} \right)\gamma  \right)}{Q_{\alpha} \left(\lambda +i\frac{\alpha}{2} \gamma  \right)}
\frac{Q_{\alpha+1} \left(\lambda +i\left(\frac{\alpha -1}{2} \right)\gamma  \right)}{Q_{\alpha+1} \left(\lambda +i\left(\frac{\alpha +1}{2} \right)\gamma  \right)}
\;\;\;\;\;\;\;\;\;\;\;\;\;\;\;\;\;\;\;\;\;\;\;\;\; \mbox{for} \;\; \alpha = 1,\dots ,m-1 \neq m+n-2 \cr
\frac{Q_{\alpha} \left(\lambda +i\left(\frac{\alpha -2}{2} \right)\gamma  \right)}{Q_{\alpha} \left(\lambda +i\frac{\alpha}{2} \gamma  \right)}
\frac{Q_{\alpha+1} \left(\lambda +i\left(\frac{\alpha +1}{2} \right)\gamma  \right)}{Q_{\alpha+1} \left(\lambda +i\left(\frac{\alpha -1}{2} \right)\gamma  \right)}
\;\;\;\;\;\;\;\;\;\;\;\;\;\;\;\;\;\;\;\;\;\;\;\;\; \mbox{for} \;\; \alpha = m \neq m+n-2 \cr
\frac{Q_{\alpha} \left(\lambda +i\left(m-1 - \frac{\alpha}{2} \right)\gamma  \right)}{Q_{\alpha} \left(\lambda +i\left( m-\frac{\alpha}{2} \right) \gamma  \right)}
\frac{Q_{\alpha+1} \left(\lambda +i\left(m + \frac{1}{2} -\frac{\alpha}{2} \right)\gamma  \right)}{Q_{\alpha+1} \left(\lambda +i\left(m-\frac{1}{2} - \frac{\alpha}{2} \right)\gamma  \right)}
\;\;\;\;\;\;\;\;\;\;\;\;\;\;\; \mbox{for} \;\; \alpha = m+1,\dots ,m+n-3 \neq m+n-2 \cr
\frac{Q_{\alpha} \left(\lambda +i\left(\frac{\alpha -2}{2} \right)\gamma  \right)}{Q_{\alpha} \left(\lambda +i\frac{\alpha}{2} \gamma  \right)}
\frac{Q_{+} \left(\lambda +i\left(\frac{\alpha +1}{2} \right)\gamma  \right)}{Q_{+} \left(\lambda +i\left(\frac{\alpha -1}{2} \right)\gamma  \right)}
\frac{Q_{-} \left(\lambda +i\left(\frac{\alpha +1}{2} \right)\gamma \right)}{Q_{-} \left(\lambda +i\left(\frac{\alpha -1}{2} \right)\gamma \right)}
\;\;\;\;\;\;\;\;\;\;\;\;\;\;\;\;\;\;\;\;\;\;\;\;\;\;\;\;\;\;\;\;\;\;\;\ \mbox{for} \;\; \alpha = m +n-2 =m \cr
\frac{Q_{\alpha} \left(\lambda +i\left(m-1 -\frac{\alpha}{2} \right)\gamma  \right)}{Q_{\alpha} \left(\lambda +i\left( m-\frac{\alpha}{2} \right) \gamma  \right)}
\frac{Q_{+} \left(\lambda +i\left(m + \frac{1}{2} -\frac{\alpha}{2} \right)\gamma  \right)}{Q_{+} \left(\lambda +i\left(m - \frac{1}{2} - \frac{\alpha}{2} \right)\gamma  \right)}
\frac{Q_{-} \left(\lambda +i\left(m + \frac{1}{2} -\frac{\alpha}{2} \right)\gamma \right)}{Q_{-} \left(\lambda +i \left(m - \frac{1}{2} - \frac{\alpha}{2} \right)\gamma \right)}
\;\;\;\;\;\;\;\;\;\;\;\;\;\;\;\;\;\;\;\;\; \mbox{for} \;\; \alpha = m+n-2 \neq m \cr
\frac{Q_{+} \left(\lambda +i\left(m + 1 -\frac{\alpha}{2} \right)\gamma  \right)}{Q_{+} \left(\lambda +i \left(m  - \frac{\alpha}{2} \right)\gamma  \right)}
\frac{Q_{-} \left(\lambda +i\left(m - 1 -\frac{\alpha}{2} \right)\gamma \right)}{Q_{-} \left(\lambda +i \left(m - \frac{\alpha}{2} \right)\gamma \right)}
\;\;\;\;\;\;\;\;\;\;\;\;\;\;\;\;\;\;\;\;\;\;\;\;\;\;\;\;\;\;\;\;\;\;\;\;\;\;\;\;\;\;\;\;\;\;\; \mbox{for} \;\; \alpha = m+n-1 \cr
G_{\alpha - (m+n-1)}(-i(m-n+1)\gamma -\lambda |- \{ \lambda_{j}^{(l_{\beta})} \}) \;\;\;\;\;\;\;\;\;\;\;\;\;\;\;
\;\;\;\; \mbox{for} \;\; \alpha = m+n,\dots ,2m+2n-2 \cr } \nonumber
\end{eqnarray}

\vskip 1.0cm
\begin{itemize}
\item $U_{q}[osp(2n+1|2m)^{(1)}]$:
\end{itemize}

\begin{eqnarray}
\Lambda^{(l_{0})} (\lambda) &=& (-1)^{p^{(l_{0})}_{1}} \left[ (-1)^{p^{(l_{0})}_{1}} a_{1}^{(l_{0})}(\lambda) \right]^{L}
\left[ \frac{Q_{1}\left( \lambda - i\frac{\gamma}{2} \right)}{Q_{1}\left( \lambda + i\frac{\gamma}{2} \right)}\right]^{2 p^{(l_{0})}_{1} -1} \nonumber \\
&+& (-1)^{p^{(l_{0})}_{N_{0}}} \left[ (-1)^{p^{(l_{0})}_{N_{0}}} d_{N_{0},N_{0}}^{(l_{0})}(\lambda) \right]^{L}
\left[ \frac{Q_{1}\left( \lambda + i\left(m-n +1\right)\gamma  \right)}{Q_{1}\left( \lambda + i\left(m-n \right)\gamma  \right)}\right]^{2 p^{(l_{0})}_{1} -1} \nonumber \\
&+& \left[ b^{(l_{0})} (\lambda)  \right]^{L} \sum_{\alpha=1}^{N_{0} -2} G_{\alpha}(\lambda | \{ \lambda_{j}^{(l_{\beta})} \}) \nonumber \\
\end{eqnarray}

\begin{eqnarray}
&& G_{\alpha}(\lambda | \{ \lambda_{j}^{(l_{\beta})} \}) \nonumber \\
&& =\cases{
\frac{Q_{\alpha} \left(\lambda +i\left(\frac{\alpha +2}{2} \right)\gamma  \right)}{Q_{\alpha} \left(\lambda +i\frac{\alpha}{2} \gamma  \right)}
\frac{Q_{\alpha+1} \left(\lambda +i\left(\frac{\alpha -1}{2} \right)\gamma  \right)}{Q_{\alpha+1} \left(\lambda +i\left(\frac{\alpha +1}{2} \right)\gamma  \right)}
\;\;\;\;\;\;\;\;\;\;\;\;\;\;\;\;\;\;\;\;\;\;\;\;\;\;\;\;\;\;\; \mbox{for} \;\; \alpha = 1,\dots ,m-1 \cr
\frac{Q_{\alpha} \left(\lambda +i\left(\frac{\alpha -2}{2} \right)\gamma  \right)}{Q_{\alpha} \left(\lambda +i\frac{\alpha}{2} \gamma  \right)}
\frac{Q_{\alpha+1} \left(\lambda +i\left(\frac{\alpha +1}{2} \right)\gamma  \right)}{Q_{\alpha+1} \left(\lambda +i\left(\frac{\alpha -1}{2} \right)\gamma  \right)}
\;\;\;\;\;\;\;\;\;\;\;\;\;\;\;\;\;\;\;\;\;\;\;\;\;\;\;\;\;\;\; \mbox{for} \;\; \alpha = m  \cr
\frac{Q_{\alpha} \left(\lambda +i\left(m-1 -\frac{\alpha}{2} \right)\gamma  \right)}{Q_{\alpha} \left(\lambda +i\left( m-\frac{\alpha}{2} \right) \gamma  \right)}
\frac{Q_{\alpha+1} \left(\lambda +i\left(m + \frac{1}{2} -\frac{\alpha}{2} \right)\gamma  \right)}{Q_{\alpha+1} \left(\lambda +i\left(m-\frac{1}{2} - \frac{\alpha}{2} \right)\gamma  \right)}
\;\;\;\;\;\;\;\;\;\;\;\;\;\;\;\;\;\;\;\;\; \mbox{for} \;\; \alpha = m+1,\dots ,m+n-1 \neq m+n \cr
\frac{Q_{\alpha} \left(\lambda +i\left(\frac{m-n -1}{2} \right)\gamma  \right)}{Q_{\alpha} \left(\lambda +i\left( \frac{m-n+1}{2} \right) \gamma \right)}
\frac{Q_{\alpha} \left(\lambda +i\left(\frac{m-n +2}{2} \right)\gamma  \right)}{Q_{\alpha} \left(\lambda +i\left(\frac{m-n}{2} \right)\gamma  \right)}
\;\;\;\;\;\;\;\;\;\;\;\;\;\;\;\;\;\;\;\;\;\;\;\;\;\;\ \mbox{for} \;\; \alpha = m+n \cr
G_{\alpha - (m+n)}(- i(m-n+\frac{1}{2})\gamma -\lambda |- \{ \lambda_{j}^{(l_{\beta})} \}) 
\;\;\;\;\;\;\;\;\;\;\;\; \mbox{for} \;\; \alpha = m+n+1,\dots ,2m+2n-1 \cr } \nonumber
\end{eqnarray}

\vskip 1.0cm
\begin{itemize}
\item $U_{q}[osp(2n|2m)^{(2)}]$:
\end{itemize}

\begin{eqnarray}
\label{grupof}
\Lambda^{(l_{0})} (\lambda) &=& (-1)^{p^{(l_{0})}_{1}} \left[ (-1)^{p^{(l_{0})}_{1}} a_{1}^{(l_{0})}(\lambda) \right]^{L}
\left[ \frac{Q_{1}\left( \lambda - i\frac{\gamma}{2} \right)}{Q_{1}\left( \lambda + i\frac{\gamma}{2} \right)}\right]^{2 p^{(l_{0})}_{1} -1} \nonumber \\
&+& (-1)^{p^{(l_{0})}_{N_{0}}} \left[ (-1)^{p^{(l_{0})}_{N_{0}}} d_{N_{0},N_{0}}^{(l_{0})}(\lambda) \right]^{L}
\left[ \frac{Q_{1}\left( \lambda + i\left(m-n-\frac{1}{2} \right)\gamma \right)}{Q_{1}\left( \lambda + i\left(m-n-\frac{3}{2} \right)\gamma \right)}\right]^{2 p^{(l_{0})}_{1} -1} \nonumber \\
&+& \left[ b^{(l_{0})} (\lambda)  \right]^{L} \sum_{\alpha=1}^{N_{0} -2} G_{\alpha}(\lambda | \{ \lambda_{j}^{(l_{\beta})} \}) \nonumber \\
\end{eqnarray}
\begin{eqnarray}
&& G_{\alpha}(\lambda | \{ \lambda_{j}^{(l_{\beta})} \}) \nonumber \\
&& =\cases{
\frac{Q_{\alpha} \left(\lambda +i\left(\frac{\alpha +2}{2} \right)\gamma  \right)}{Q_{\alpha} \left(\lambda +i\frac{\alpha}{2} \gamma  \right)}
\frac{Q_{\alpha+1} \left(\lambda +i\left(\frac{\alpha -1}{2} \right)\gamma  \right)}{Q_{\alpha+1} \left(\lambda +i\left(\frac{\alpha +1}{2} \right)\gamma  \right)}
\;\;\;\;\;\;\;\;\;\;\;\;\;\;\;\;\;\;\;\;\;\; \mbox{for} \;\; \alpha = 1,\dots ,m-1 \cr
\frac{Q_{\alpha} \left(\lambda +i\left(\frac{\alpha -2}{2} \right)\gamma  \right)}{Q_{\alpha} \left(\lambda +i\frac{\alpha}{2} \gamma  \right)}
\frac{Q_{\alpha+1} \left(\lambda +i\left(\frac{\alpha +1}{2} \right)\gamma  \right)}{Q_{\alpha+1} \left(\lambda +i\left(\frac{\alpha -1}{2} \right)\gamma  \right)}
\;\;\;\;\;\;\;\;\;\;\;\;\;\;\;\;\;\;\;\;\;\; \mbox{for} \;\; \alpha = m \neq m+n-1 \cr
\frac{Q_{\alpha} \left(\lambda +i\left(m-1- \frac{\alpha}{2} \right)\gamma  \right)}{Q_{\alpha} \left(\lambda +i\left( m-\frac{\alpha}{2} \right) \gamma  \right)}
\frac{Q_{\alpha+1} \left(\lambda +i\left(m + \frac{1}{2} -\frac{\alpha}{2} \right)\gamma  \right)}{Q_{\alpha+1} \left(\lambda +i\left(m-\frac{1}{2} - \frac{\alpha}{2} \right)\gamma  \right)}
\;\;\;\;\;\;\;\;\;\;\;\; \mbox{for} \;\; \alpha = m+1,\dots ,m+n-2 \neq m+n-1 \cr
\frac{Q_{\alpha} \left(\lambda +i\left(m-1- \frac{\alpha}{2} \right)\gamma  \right)}{Q_{\alpha} \left(\lambda +i\left( m-\frac{\alpha}{2} \right) \gamma  \right)}
\frac{Q_{\alpha+1} \left(\lambda +i\left(m + 1 -\frac{\alpha}{2} \right)\gamma  \right)}{Q_{\alpha+1} \left(\lambda +i\left(m-1 - \frac{\alpha}{2} \right)\gamma  \right)}
\;\;\;\;\;\;\;\;\;\;\;\;\;\;\;\;\;\;\;\;\;\;\;\;\; \mbox{for} \;\; \alpha = m+n-1 \neq m \cr
\frac{Q_{\alpha} \left(\lambda +i\left(\frac{\alpha -2}{2} \right)\gamma  \right)}{Q_{\alpha} \left(\lambda +i\frac{\alpha}{2} \gamma  \right)}
\frac{Q_{\alpha+1} \left(\lambda +i\left(\frac{\alpha +2}{2} \right)\gamma  \right)}{Q_{\alpha+1} \left(\lambda +i\left(\frac{\alpha -2}{2} \right)\gamma  \right)}
\;\;\;\;\;\;\;\;\;\;\;\;\;\;\;\;\;\;\;\;\;\;\;\;\;\;\;\;\;\;\;\;\;\; \mbox{for} \;\;\alpha = m+n-1 = m \cr
G_{\alpha - (m+n-1)}(-i(m-n-1)\gamma -\lambda |- \{ \lambda_{j}^{(l_{\beta})} \}) \;\;\;\;\;\;\;\;\;\;\;\; \mbox{for} \;\; \alpha = m+n,\dots ,2m+2n-2 \cr }. \nonumber
\end{eqnarray}

Before proceeding with the Bethe ansatz equations, we note that in the
expressions  (\ref{grupo1}-\ref{grupof}) we have  performed the shifts  $\{ \lambda_j^{(l_{\alpha})} \} \rightarrow
\{ \lambda_j^{(l_{\alpha})} \}-\delta^{(l_{\alpha})}$ in order to bring the final results
in a more symmetrical way.  In Table 3 we show the values for the  displacements
$\delta^{(l_{\alpha})}$.  The same procedure described above for the eigenvalues also
works for determining the Bethe ansatz equations for the shifted rapidities. We begin with
Eq.(\ref{bar}), each step of the nesting is disentangled with the help of the relation (\ref{beterec}) and when
we reach the last step $l_f$ we use the Bethe ansatz results exhibited in appendix C.
It turns out that the Bethe ansatz equations for these vertex models are given by

\begin{itemize}
\item $U_{q}[sl(2n+1|2m)^{(2)}]$:
\end{itemize}

\begin{eqnarray}
\prod_{i=1}^{m_{l_{\alpha -1}}} \frac{\sinh{\left(\lambda_{j}^{(l_{\alpha})} -\lambda_{i}^{(l_{\alpha-1})} +i\frac{\gamma}{2} \right)}}{\sinh{\left(\lambda_{j}^{(l_{\alpha})} -\lambda_{i}^{(l_{\alpha-1})} -i\frac{\gamma}{2} \right)}}&=&
\prod_{i\neq j}^{m_{l_{\alpha}}} \frac{\sinh{\left(\lambda_{j}^{(l_{\alpha})} -\lambda_{i}^{(l_{\alpha})} +i\gamma \right)}}{\sinh{\left(\lambda_{j}^{(l_{\alpha})} -\lambda_{i}^{(l_{\alpha})} -i\gamma \right)}}
\prod_{i=1}^{m_{l_{\alpha +1}}} \frac{\sinh{\left(\lambda_{i}^{(l_{\alpha+1})} -\lambda_{j}^{(l_{\alpha})} + i\frac{\gamma}{2} \right)}}{\sinh{\left(\lambda_{i}^{(l_{\alpha+1})} -\lambda_{j}^{(l_{\alpha})} -  i\frac{\gamma}{2} \right)}} \nonumber \\
&& \;\;\;\;\;\;\;\;\;\;\;\;\;\;\;\;\;\;\;\;\;\;\;\;\;\;\;\;\;\;\;\;\;\;\;\;\;\;\;\;\;\;\;\;\;\;\;\;\;\;\;\;\;\;\;\;\;\;\;\;\;\;\;\; \alpha= 1,\dots , m-1  \nonumber \\
\prod_{i=1}^{m_{l_{\alpha -1}}} \frac{\sinh{\left(\lambda_{j}^{(l_{\alpha})} -\lambda_{i}^{(l_{\alpha-1})} +i\frac{\gamma}{2} \right)}}{\sinh{\left(\lambda_{j}^{(l_{\alpha})} -\lambda_{i}^{(l_{\alpha-1})} -i\frac{\gamma}{2} \right)}}&=&
\prod_{i=1}^{m_{l_{\alpha +1}}} \frac{\sinh{\left(\lambda_{i}^{(l_{\alpha+1})} -\lambda_{j}^{(l_{\alpha})} - i\frac{\gamma}{2} \right)}}{\sinh{\left(\lambda_{i}^{(l_{\alpha+1})} -\lambda_{j}^{(l_{\alpha})} + i\frac{\gamma}{2} \right)}}
\;\;\;\;\;\;\;\;\;\;\;\;\;\;\;\;\;\;\;\;\;\;\;\;\;\;\;\;\;\;\;\;\;\; \alpha=m  \nonumber \\
\prod_{i=1}^{m_{l_{\alpha -1}}} \frac{\sinh{\left(\lambda_{j}^{(l_{\alpha})} -\lambda_{i}^{(l_{\alpha-1})} -i\frac{\gamma}{2} \right)}}{\sinh{\left(\lambda_{j}^{(l_{\alpha})} -\lambda_{i}^{(l_{\alpha-1})} +i\frac{\gamma}{2} \right)}}&=&
\prod_{i\neq j}^{m_{l_{\alpha}}} \frac{\sinh{\left(\lambda_{j}^{(l_{\alpha})} -\lambda_{i}^{(l_{\alpha})} -i\gamma \right)}}{\sinh{\left(\lambda_{j}^{(l_{\alpha})} -\lambda_{i}^{(l_{\alpha})} +i\gamma \right)}}
\prod_{i=1}^{m_{l_{\alpha +1}}} \frac{\sinh{\left(\lambda_{i}^{(l_{\alpha+1})} -\lambda_{j}^{(l_{\alpha})} - i\frac{\gamma}{2} \right)}}{\sinh{\left(\lambda_{i}^{l_{(\alpha+1})} -\lambda_{j}^{(l_{\alpha})} + i\frac{\gamma}{2} \right)}} \nonumber \\
&& \;\;\;\;\;\;\;\;\;\;\;\;\;\;\;\;\;\;\;\;\;\;\;\;\;\;\;\;\;\; \alpha= m+1,\dots , m+n-1 \neq m+n \neq m \nonumber \\
\prod_{i=1}^{m_{l_{\alpha -1}}} \frac{\sinh{\left(\lambda_{j}^{(l_{\alpha})} -\lambda_{i}^{(l_{\alpha-1})} -i\frac{\gamma}{2} \right)}}{\sinh{\left(\lambda_{j}^{(l_{\alpha})} -\lambda_{i}^{(l_{\alpha-1})} +i\frac{\gamma}{2} \right)}}&=&
\prod_{i\neq j}^{m_{l_{\alpha}}} \frac{\sinh{\left(\lambda_{j}^{(l_{\alpha})} -\lambda_{i}^{(l_{\alpha})} -i\gamma \right)}}{\sinh{\left(\lambda_{j}^{(l_{\alpha})} -\lambda_{i}^{(l_{\alpha})} +i\gamma \right)}}
\frac{\cosh{\left(\lambda_{j}^{(l_{\alpha})} -\lambda_{i}^{(l_{\alpha})} +i\frac{\gamma}{2} \right)}}{\cosh{\left(\lambda_{j}^{(l_{\alpha})} -\lambda_{i}^{(l_{\alpha})} - i\frac{\gamma}{2} \right)}} \nonumber \\
&& \;\;\;\;\;\;\;\;\;\;\;\;\;\;\;\;\;\;\;\;\;\;\;\;\;\;\;\;\;\;\;\;\;\;\;\;\;\;\;\;\;\;\;\;\;\;\;\;\;\;\;\;\;\;\;\;\;\;\;\;\;\;\;\;\;\;\;\;\;\;\;\;\;\; \alpha=m+n \nonumber \\
\end{eqnarray}

\begin{itemize}
\item $U_{q}[sl(2n|2m)^{(2)}]$:
\end{itemize}

\begin{eqnarray}
\prod_{i=1}^{m_{l_{\alpha -1}}} \frac{\sinh{\left(\lambda_{j}^{(l_{\alpha})} -\lambda_{i}^{(l_{\alpha-1})} +i\frac{\gamma}{2} \right)}}{\sinh{\left(\lambda_{j}^{(l_{\alpha})} -\lambda_{i}^{(l_{\alpha-1})} -i\frac{\gamma}{2} \right)}}&=&
\prod_{i\neq j}^{m_{l_{\alpha}}} \frac{\sinh{\left(\lambda_{j}^{(l_{\alpha})} -\lambda_{i}^{(l_{\alpha})} +i\gamma \right)}}{\sinh{\left(\lambda_{j}^{(l_{\alpha})} -\lambda_{i}^{(l_{\alpha})} -i\gamma \right)}}
\prod_{i=1}^{m_{l_{\alpha +1}}} \frac{\sinh{\left(\lambda_{i}^{(l_{\alpha+1})} -\lambda_{j}^{(l_{\alpha})} + i\frac{\gamma}{2} \right)}}{\sinh{\left(\lambda_{i}^{(l_{\alpha+1})} -\lambda_{j}^{(l_{\alpha})} -  i\frac{\gamma}{2} \right)}} \nonumber \\
&&  \;\;\;\;\;\;\;\;\;\;\;\;\;\;\;\;\;\;\;\;\;\;\;\;\;\;\;\;\;\;\;\;\;\;\;\;\;\;\;\;\;\;\;\;\;\;\;\;\;\;\;\;\;\;\;\;\;\;\;\;\;\;\; \alpha= 1,\dots , m-1  \nonumber \\
\prod_{i=1}^{m_{l_{\alpha -1}}} \frac{\sinh{\left(\lambda_{j}^{(l_{\alpha})} -\lambda_{i}^{(l_{\alpha-1})} +i\frac{\gamma}{2} \right)}}{\sinh{\left(\lambda_{j}^{(l_{\alpha})} -\lambda_{i}^{(l_{\alpha-1})} -i\frac{\gamma}{2} \right)}}&=&
\prod_{i=1}^{m_{l_{\alpha +1}}} \frac{\sinh{\left[g_{\alpha} \left(\lambda_{i}^{(l_{\alpha+1})} -\lambda_{j}^{(l_{\alpha})} -  i\frac{\gamma}{2} \right) \right]}}{\sinh{\left[ g_{\alpha} \left( \lambda_{i}^{(l_{\alpha+1})} -\lambda_{j}^{(l_{\alpha})} +  i\frac{\gamma}{2} \right) \right]}}
\;\;\;\;\;\;\;\;\;\;\;\;\;\;\;\;\;\;\;\;\;\;\;\;\; \alpha=m  \nonumber \\
\prod_{i=1}^{m_{l_{\alpha -1}}} \frac{\sinh{\left(\lambda_{j}^{(l_{\alpha})} -\lambda_{i}^{(l_{\alpha-1})} -i\frac{\gamma}{2} \right)}}{\sinh{\left(\lambda_{j}^{(l_{\alpha})} -\lambda_{i}^{l_{(\alpha-1})} +i\frac{\gamma}{2} \right)}}&=&
\prod_{i\neq j}^{m_{l_{\alpha}}} \frac{\sinh{\left(\lambda_{j}^{(l_{\alpha})} -\lambda_{i}^{(l_{\alpha})} -i\gamma \right)}}{\sinh{\left(\lambda_{j}^{(l_{\alpha})} -\lambda_{i}^{(l_{\alpha})} +i\gamma \right)}}
\prod_{i=1}^{m_{l_{\alpha +1}}} \frac{\sinh{\left[g_{\alpha} \left(\lambda_{i}^{(l_{\alpha+1})} -\lambda_{j}^{(l_{\alpha})} -  i\frac{\gamma}{2} \right) \right]}}{\sinh{\left[ g_{\alpha} \left( \lambda_{i}^{(l_{\alpha+1})} -\lambda_{j}^{(l_{\alpha})} + i \frac{\gamma}{2} \right) \right]}} \nonumber \\
&& \;\;\;\;\;\;\;\;\;\;\;\;\;\;\;\;\;\;\;\;\;\;\;\;\;\;\;\; \alpha= m+1,\dots , m+n-1 \neq m+n \neq m \nonumber \\
\prod_{i=1}^{m_{l_{\alpha -1}}} \frac{\sinh{\left[ 2 \left( \lambda_{j}^{(l_{\alpha})} -\lambda_{i}^{(l_{\alpha-1})} -i\frac{\gamma}{2} \right) \right]}}{\sinh{\left[ 2 \left( \lambda_{j}^{(l_{\alpha})} -\lambda_{i}^{(l_{\alpha-1})} +i\frac{\gamma}{2} \right) \right]}}&=&
\prod_{i\neq j}^{m_{l_{\alpha}}} \frac{\sinh{\left[ 2 \left( \lambda_{j}^{(l_{\alpha})} -\lambda_{i}^{(l_{\alpha})} -i\gamma \right) \right]}}{\sinh{\left[ 2 \left( \lambda_{j}^{(l_{\alpha})} -\lambda_{i}^{(l_{\alpha})} +i\gamma \right) \right]}}
\;\;\;\;\;\;\;\;\;\;\;\;\;\;\;\;\;\;\;\;\;\;\;\;\;\; \alpha=m+n \nonumber \\
\end{eqnarray}

\begin{itemize}
\item $U_{q}[osp(2n|2m)^{(1)}]$:
\end{itemize}

\begin{eqnarray}
\prod_{i=1}^{m_{l_{\alpha -1}}} \frac{\sinh{\left(\lambda_{j}^{(l_{\alpha})} -\lambda_{i}^{(l_{\alpha-1})} +i\frac{\gamma}{2} \right)}}{\sinh{\left(\lambda_{j}^{(l_{\alpha})} -\lambda_{i}^{(l_{\alpha-1})} -i\frac{\gamma}{2} \right)}}&=&
\prod_{i\neq j}^{m_{l_{\alpha}}} \frac{\sinh{\left(\lambda_{j}^{(l_{\alpha})} -\lambda_{i}^{(l_{\alpha})} +i\gamma \right)}}{\sinh{\left(\lambda_{j}^{(l_{\alpha})} -\lambda_{i}^{(l_{\alpha})} -i\gamma \right)}}
\prod_{i=1}^{m_{l_{\alpha +1}}} \frac{\sinh{\left(\lambda_{i}^{l_{(\alpha+1})} -\lambda_{j}^{(l_{\alpha})} + i\frac{\gamma}{2} \right)}}{\sinh{\left(\lambda_{i}^{(l_{\alpha+1})} -\lambda_{j}^{(l_{\alpha})} - i \frac{\gamma}{2} \right)}} \nonumber \\
&& \;\;\;\;\;\;\;\;\;\;\;\;\;\;\;\;\;\;\;\;\;\;\;\;\;\;\;\;\;\;\;\;\;\;\;\;\;\;\;\;\;\;\;\;\;\;\;\;\;\;\;\;\;\;\;\;\;\;\;\;\;\;\;\; \alpha= 1,\dots , m-1  \nonumber \\
\prod_{i=1}^{m_{l_{\alpha -1}}} \frac{\sinh{\left(\lambda_{j}^{(l_{\alpha})} -\lambda_{i}^{(l_{\alpha-1})} +i\frac{\gamma}{2} \right)}}{\sinh{\left(\lambda_{j}^{(l_{\alpha})} -\lambda_{i}^{(l_{\alpha-1})} -i\frac{\gamma}{2} \right)}}&=&
\prod_{i=1}^{m_{l_{\alpha +1}}} \frac{\sinh{\left(\lambda_{i}^{(l_{\alpha+1})} -\lambda_{j}^{(l_{\alpha})} - i\frac{\gamma}{2} \right)}}{\sinh{\left(\lambda_{i}^{(l_{\alpha+1})} -\lambda_{j}^{(l_{\alpha})} + i\frac{\gamma}{2} \right)}}
\;\;\;\;\;\;\;\;\;\;\;\;\;\;\;\;\;\;\;\;\;\;\;\;\;\;\;\;\;\;\;\;\; \alpha=m  \nonumber \\
\prod_{i=1}^{m_{l_{\alpha -1}}} \frac{\sinh{\left(\lambda_{j}^{(l_{\alpha})} -\lambda_{i}^{(l_{\alpha-1})} -i\frac{\gamma}{2} \right)}}{\sinh{\left(\lambda_{j}^{(l_{\alpha})} -\lambda_{i}^{(l_{\alpha-1})} +i\frac{\gamma}{2} \right)}}&=&
\prod_{i\neq j}^{m_{l_{\alpha}}} \frac{\sinh{\left(\lambda_{j}^{(l_{\alpha})} -\lambda_{i}^{(l_{\alpha})} -i\gamma \right)}}{\sinh{\left(\lambda_{j}^{(l_{\alpha})} -\lambda_{i}^{(l_{\alpha})} +i\gamma \right)}}
\prod_{i=1}^{m_{l_{\alpha +1}}} \frac{\sinh{\left(\lambda_{i}^{(l_{\alpha+1})} -\lambda_{j}^{(l_{\alpha})} - i\frac{\gamma}{2} \right)}}{\sinh{\left(\lambda_{i}^{(l_{\alpha+1})} -\lambda_{j}^{(l_{\alpha})} + i\frac{\gamma}{2} \right)}} \nonumber \\
&& \;\;\;\;\;\;\;\;\;\;\;\;\;\;\;\;\;\;\;\;\;\;\; 
\alpha= m+1,\dots , m+n-3 \neq m+n-2 \neq m \nonumber \\
\prod_{i=1}^{m_{l_{\alpha -1}}} \frac{\sinh{\left(\lambda_{j}^{(l_{\alpha})} -\lambda_{i}^{(l_{\alpha-1})} -i\frac{\gamma}{2} \right)}}{\sinh{\left(\lambda_{j}^{(l_{\alpha})} -\lambda_{i}^{(l_{\alpha-1})} +i\frac{\gamma}{2} \right)}}&=&
\prod_{i\neq j}^{m_{l_{\alpha}}} \frac{\sinh{\left(\lambda_{j}^{(l_{\alpha})} -\lambda_{i}^{(l_{\alpha})} -i\gamma \right)}}{\sinh{\left(\lambda_{j}^{(l_{\alpha})} -\lambda_{i}^{(l_{\alpha})} +i\gamma \right)}}\nonumber \\
&\times& \prod_{i=1}^{m_{l_{+}}} \frac{\sinh{\left(\lambda_{i}^{(l_{+})} -\lambda_{j}^{(l_{\alpha})} - i\frac{\gamma}{2} \right)}}{\sinh{\left(\lambda_{i}^{(l_{+})} -\lambda_{j}^{(l_{\alpha})} + i\frac{\gamma}{2} \right)}}
\prod_{i=1}^{m_{l_{-}}} \frac{\sinh{\left(\lambda_{i}^{(l_{-})} -\lambda_{j}^{(l_{\alpha})} - i\frac{\gamma}{2} \right)}}{\sinh{\left(\lambda_{i}^{(l_{-})} -\lambda_{j}^{(l_{\alpha})} +i \frac{\gamma}{2} \right)}} \nonumber \\
&& \;\;\;\;\;\;\;\;\;\;\;\;\;\;\;\;\;\;\;\;\;\;\;\;\;\;\;\;\;\;\;\;\;\;\;\;\;\;\;\;\;\;\;\;\;\;\;\;\;\;\;\;\;\;\;\;\;\;\;\;\;\;\;\;\;\;\;\; \alpha= m+n-2 \nonumber \\
\prod_{i=1}^{m_{l_{\alpha -1}}} \frac{\sinh{\left(\lambda_{j}^{(l_{\pm})} -\lambda_{i}^{(l_{\alpha-1})} -i\frac{\gamma}{2} \right)}}{\sinh{\left(\lambda_{j}^{(l_{\pm})} -\lambda_{i}^{l_{(\alpha-1})} +i\frac{\gamma}{2} \right)}}&=&
\prod_{i\neq j}^{m_{l_{\pm}}} \frac{\sinh{\left(\lambda_{j}^{(l_{\pm})} -\lambda_{i}^{(l_{\pm})} -i\gamma \right)}}{\sinh{\left(\lambda_{j}^{(l_{\pm})} -\lambda_{i}^{(l_{\pm})} +i\gamma \right)}}
\;\;\;\;\;\;\;\;\;\;\;\;\;\;\;\;\;\;\;\;\;\;\;\;\;\; \alpha=m+n-1  \nonumber \\
\end{eqnarray}

\vskip 2.0cm
\begin{itemize}
\item $U_{q}[osp(2n+1|2m)^{(1)}]$:
\end{itemize}
\begin{eqnarray}
\prod_{i=1}^{m_{l_{\alpha -1}}} \frac{\sinh{\left(\lambda_{j}^{(l_{\alpha})} -\lambda_{i}^{(l_{\alpha-1})} +i\frac{\gamma}{2} \right)}}{\sinh{\left(\lambda_{j}^{(l_{\alpha})} -\lambda_{i}^{(l_{\alpha-1})} -i\frac{\gamma}{2} \right)}}&=&
\prod_{i\neq j}^{m_{l_{\alpha}}} \frac{\sinh{\left(\lambda_{j}^{(l_{\alpha})} -\lambda_{i}^{(l_{\alpha})} +i\gamma \right)}}{\sinh{\left(\lambda_{j}^{(l_{\alpha})} -\lambda_{i}^{(l_{\alpha})} -i\gamma \right]}}
\prod_{i=1}^{m_{l_{\alpha +1}}} \frac{\sinh{\left(\lambda_{i}^{(l_{\alpha+1})} -\lambda_{j}^{(l_{\alpha})} + i\frac{\gamma}{2} \right)}}{\sinh{\left(\lambda_{i}^{l_{\alpha+1}} -\lambda_{j}^{(l_{\alpha})} -  i\frac{\gamma}{2} \right)}} \nonumber \\
&&  \;\;\;\;\;\;\;\;\;\;\;\;\;\;\;\;\;\;\;\;\;\;\;\;\;\;\;\;\;\;\;\;\;\;\;\;\;\;\;\;\;\;\;\;\;\;\;\;\;\;\;\;\;\;\;\;\;\;\;\;\;\;\;\; \alpha= 1,\dots , m-1  \nonumber \\
\prod_{i=1}^{m_{l_{\alpha -1}}} \frac{\sinh{\left(\lambda_{j}^{(l_{\alpha})} -\lambda_{i}^{(l_{\alpha-1})} +i\frac{\gamma}{2} \right)}}{\sinh{\left(\lambda_{j}^{(l_{\alpha})} -\lambda_{i}^{(l_{\alpha-1})} -i\frac{\gamma}{2} \right)}}&=&
\prod_{i=1}^{m_{l_{\alpha +1}}} \frac{\sinh{\left(\lambda_{i}^{(l_{\alpha+1})} -\lambda_{j}^{(l_{\alpha})} -i\frac{\gamma}{2} \right)}}{\sinh{\left(\lambda_{i}^{(l_{\alpha+1})} -\lambda_{j}^{(l_{\alpha})} + i\frac{\gamma}{2} \right)}}
\;\;\;\;\;\;\;\;\;\;\;\;\;\;\;\;\;\;\;\;\;\;\;\;\;\;\;\;\;\;\;\;\; \alpha=m  \nonumber \\
\prod_{i=1}^{m_{l_{\alpha -1}}} \frac{\sinh{\left(\lambda_{j}^{(l_{\alpha})} -\lambda_{i}^{(l_{\alpha-1})} -i\frac{\gamma}{2} \right)}}{\sinh{\left(\lambda_{j}^{(l_{\alpha})} -\lambda_{i}^{(l_{\alpha-1})} +i\frac{\gamma}{2} \right)}}&=&
\prod_{i\neq j}^{m_{l_{\alpha}}} \frac{\sinh{\left(\lambda_{j}^{(l_{\alpha})} -\lambda_{i}^{(l_{\alpha})} -i\gamma \right)}}{\sinh{\left(\lambda_{j}^{(l_{\alpha})} -\lambda_{i}^{(l_{\alpha})} +i\gamma \right)}}
\prod_{i=1}^{m_{l_{\alpha +1}}} \frac{\sinh{\left(\lambda_{i}^{(l_{\alpha+1})} -\lambda_{j}^{(l_{\alpha})} - i\frac{\gamma}{2} \right)}}{\sinh{\left(\lambda_{i}^{(l_{\alpha+1})} -\lambda_{j}^{(l_{\alpha})} + i\frac{\gamma}{2} \right)}} \nonumber \\
&& \;\;\;\;\;\;\;\;\;\;\;\;\;\;\;\;\;\;\;\;\;\;\;\;\;\;\;\;\; 
\alpha= m+1,\dots , m+n-1 \neq m+n \neq m \nonumber \\
\prod_{i=1}^{m_{l_{\alpha -1}}} \frac{\sinh{\left(\lambda_{j}^{(l_{\alpha})} -\lambda_{i}^{(l_{\alpha-1})} -i\frac{\gamma}{2} \right)}}{\sinh{\left(\lambda_{j}^{(l_{\alpha})} -\lambda_{i}^{(l_{\alpha-1})} +i\frac{\gamma}{2} \right)}}&=&
\prod_{i\neq j}^{m_{l_{\alpha}}} \frac{\sinh{\left(\lambda_{j}^{(l_{\alpha})} -\lambda_{i}^{(l_{\alpha})} -i\frac{\gamma}{2} \right)}}{\sinh{\left(\lambda_{j}^{(l_{\alpha})} -\lambda_{i}^{(l_{\alpha})} + i\frac{\gamma}{2} \right)}}
\;\;\;\;\;\;\;\;\;\;\;\;\;\;\;\;\;\;\;\;\;\;\;\;\;\;\;\;\;\;\;\; \alpha=m+n \nonumber \\
\end{eqnarray}

\vskip 1.0cm
\begin{itemize}
\item $U_{q}[osp(2n|2m)^{(2)}]$:
\end{itemize}

\begin{eqnarray}
\prod_{i=1}^{m_{l_{\alpha -1}}} \frac{\sinh{\left(\lambda_{j}^{(l_{\alpha})} -\lambda_{i}^{(l_{\alpha-1})} +i\frac{\gamma}{2} \right)}}{\sinh{\left(\lambda_{j}^{(l_{\alpha})} -\lambda_{i}^{(l_{\alpha-1})} -i\frac{\gamma}{2} \right)}}&=&
\prod_{i\neq j}^{m_{l_{\alpha}}} \frac{\sinh{\left(\lambda_{j}^{(l_{\alpha})} -\lambda_{i}^{(l_{\alpha})} +i\gamma \right)}}{\sinh{\left(\lambda_{j}^{(l_{\alpha})} -\lambda_{i}^{(l_{\alpha})} -i\gamma \right)}}
\prod_{i=1}^{m_{l_{\alpha +1}}} \frac{\sinh{\left(\lambda_{i}^{(l_{\alpha+1})} -\lambda_{j}^{(l_{\alpha})} + i\frac{\gamma}{2} \right)}}{\sinh{\left(\lambda_{i}^{(l_{\alpha+1})} -\lambda_{j}^{(l_{\alpha})} -  i\frac{\gamma}{2} \right)}} \nonumber \\
&&  \;\;\;\;\;\;\;\;\;\;\;\;\;\;\;\;\;\;\;\;\;\;\;\;\;\;\;\;\;\;\;\;\;\;\;\;\;\;\;\;\;\;\;\;\;\;\;\;\;\;\;\;\;\;\;\;\;\;\;\;\;\;\;\; \alpha= 1,\dots , m-1  \nonumber \\
\prod_{i=1}^{m_{l_{\alpha -1}}} \frac{\sinh{\left(\lambda_{j}^{(l_{\alpha})} -\lambda_{i}^{(l_{\alpha-1})} +i\frac{\gamma}{2} \right)}}{\sinh{\left(\lambda_{j}^{(l_{\alpha})} -\lambda_{i}^{(l_{\alpha-1})} -i\frac{\gamma}{2} \right)}}&=&
\prod_{i=1}^{m_{l_{\alpha +1}}} \frac{\sinh{\left(\lambda_{i}^{(l_{\alpha+1})} -\lambda_{j}^{(l_{\alpha})} - ig_{\alpha} \frac{\gamma}{2} \right)}}{\sinh{\left(\lambda_{i}^{(l_{\alpha+1})} -\lambda_{j}^{(l_{\alpha})} +i g_{\alpha} \frac{\gamma}{2} \right)}}
\;\;\;\;\;\;\;\;\;\;\;\;\;\;\;\;\;\;\;\;\;\;\;\;\;\;\;\;\; \alpha=m  \nonumber \\
\prod_{i=1}^{m_{l_{\alpha -1}}} \frac{\sinh{\left(\lambda_{j}^{(l_{\alpha})} -\lambda_{i}^{(l_{\alpha-1})} -i\frac{\gamma}{2} \right)}}{\sinh{\left(\lambda_{j}^{(l_{\alpha})} -\lambda_{i}^{(l_{\alpha-1})} +i\frac{\gamma}{2} \right)}}&=&
\prod_{i\neq j}^{m_{l_{\alpha}}} \frac{\sinh{\left(\lambda_{j}^{(l_{\alpha})} -\lambda_{i}^{(l_{\alpha})} -i\gamma \right)}}{\sinh{\left(\lambda_{j}^{(l_{\alpha})} -\lambda_{i}^{(l_{\alpha})} +i\gamma \right)}}
\prod_{i=1}^{m_{l_{\alpha +1}}} \frac{\sinh{\left(\lambda_{i}^{(l_{\alpha+1})} -\lambda_{j}^{(l_{\alpha})} - ig_{\alpha} \frac{\gamma}{2} \right)}}{\sinh{\left(\lambda_{i}^{(l_{\alpha+1})} -\lambda_{j}^{(l_{\alpha})} + ig_{\alpha} \frac{\gamma}{2} \right)}} \nonumber \\
&& \;\;\;\;\;\;\;\;\;\;\;\;\;\;\;\;\;\;\;\;\;\;\;\;\;\;\;\;\; \alpha= m+1,\dots , m+n-1 \neq m+n \neq m \nonumber \\
\prod_{i=1}^{m_{l_{\alpha -1}}} \frac{\sinh{\left(\lambda_{j}^{(l_{\alpha})} -\lambda_{i}^{(l_{\alpha-1})} -i\gamma \right)}}{\sinh{\left(\lambda_{j}^{(l_{\alpha})} -\lambda_{i}^{(l_{\alpha-1})} +i\gamma \right)}}&=&
\prod_{i\neq j}^{m_{l_{\alpha}}} \frac{\sinh{\left(\lambda_{j}^{(l_{\alpha})} -\lambda_{i}^{(l_{\alpha})} -2i\gamma \right)}}{\sinh{\left(\lambda_{j}^{(l_{\alpha})} -\lambda_{i}^{(l_{\alpha})} +2i\gamma \right)}}
\;\;\;\;\;\;\;\;\;\;\;\;\;\;\;\;\;\;\;\;\;\;\;\;\;\;\;\;\;\;\; \alpha=m+n \nonumber \\
\end{eqnarray}
where $g_{\alpha}$ has two possible values  defined by
\begin{equation}
g_{\alpha}=\cases{
2 \;\;\;\;\;\;\;\;\; \alpha = m+n-1 \cr
1 \;\;\;\;\;\;\;\;\; \mbox{otherwise} \cr }
\end{equation}

For the sake of 
completeness we have presented 
the Bethe ansatz results concerning the vertex models $U_q[sl(1|2m)^{(2)}]$ and $U_q[osp(1|2m)^{(1)}]$  
in appendix D. 
We close this section with the following remark. It is possible to verify that the systems of Bethe ansatz
equations exhibited above are the conditions of analyticity of $\Lambda(\lambda)$ as a function of
the rapidities $\{ \lambda_j^{(l_1)} \}, \dots,
\{ \lambda_j^{(l_{f+1})} \}$. This is indeed an extra check of the validity of our Bethe
ansatz results since the eigenvalue does not know a priori about the existence of such poles.

\section{Conclusion}

In this paper we have presented the $\check{R}$-matrices of the fundamental trigonometric
vertex models based on the superalgebras $sl(r|2m)^{(2)}$,
$osp(r|2m)^{(1)}$ and
$osp(2n|2m)^{(2)}$ in terms of the Weyl basis. The structure of the corresponding 
Boltzmann weights is therefore explicitly
unveiled, opening up  an opportunity to investigate the physical properties of such
vertex models from the statistical mechanics viewpoint. In fact, the transfer matrix eigenvalue
problem was formulated and solved  by a first principle algebraic framework called
quantum inverse scattering method. From our results for the transfer matrix
eigenvalues and Bethe ansatz equations one can in principle derive the free-energy thermodynamics,
the quasi-particle excitation behaviour as well as the classes  of universality governing
the criticality of gapless regimes. Furthermore, the rather universal formula we obtained 
for the eigenvectors could be useful in future computations of off-shell properties such as
form factors \cite{BAB} and correlation functions \cite{KO,MAI} of relevant operators.

This work also paves the way to undertake formal study of these vertex models with open 
boundary conditions \cite{SK}. We remark  that the $\check{R}^{(l_0)}_{ab}(\lambda)$
commutes for different values of the rapidity $\lambda$. As a consequence of that the
trivial diagonal solution of the 
reflection equation $K_{-}(\lambda)=Id$ and $K_{+}(\lambda)=V^{st}V$ \cite{NEPO} does hold for
all these vertex models. It seems an interesting problem to classify the solutions of the reflexion
equation for such models, extending the recent efforts made in the case of super-Yangian 
$R$-matrices \cite{AR} by employing for instance the technique developed in ref.\cite{LS}.

Finally, we observe that the vertex models discussed in this paper share a common algebraic
structure  denominated braid-monoid algebra \cite{WAD}.  We hope that this property will help
us to improve our understanding of such systems and to provide new insights into other related
problems.  One of them would be the explicit formulae for the $\check{R}$-matrices in an
arbitrary grading structure, which we  plan to study in a future publication.

\section*{Acknowledgements}
The authors thanks Fapesp (Funda\c c\~ao de Amparo \`a Pesquisa
do Estado de S\~ao Paulo) and
the Brazilian Research Council-CNPq  for financial support.

\section*{\bf Appendix A : The crossing symmetry}
\setcounter{equation}{0}
\renewcommand{\theequation}{A.\arabic{equation}}

The purpose of this appendix is to present
the explicit expressions for the
crossing parameter $\eta$, the normalization function $\rho(\lambda)$ and the crossing matrix $V$.
The crossing parameter $\eta$ is better written in terms of
the anisotropy $\gamma$ such that $q=e^{i\gamma}$ and it turns out that
$\eta=i(m-n+1)\gamma ,i\frac{\pi}{2} +i(m-n-\frac{1}{2})\gamma,i\frac{\pi}{2} +i(m-n)\gamma ,
i(m-n+\frac{1}{2})\gamma,i(m-n-1)\gamma$
for the superalgebras
$U_{q}[osp(2n|2m)^{(1)}]$, $U_{q}[sl(2n+1|2m)^{(2)}]$, $U_{q}[sl(2n|2m)^{(2)}]$,
$U_{q}[osp(2n+1|2m)^{(1)}]$ and $U_{q}[osp(2n|2m)^{(2)}]$, respectively.
The normalization function is given by
\begin{equation}
\rho(\lambda) =q(e^{2 \lambda} -1)(e^{2 \lambda} -\zeta^{(l_{0})})
\end{equation}

The only non-null entries of the matrix $V$ are the anti-diagonal elements $V_{\alpha {\alpha}^{'}}$.
Up to an arbitrary normalization and for each superalgebra discussed here they are
\begin{itemize}
\item $U_{q}\left[ osp(2n|2m)^{(1)} \right]$ and $U_{q}\left[sl(2n|2m)^{(2)} \right]$:
\begin{eqnarray}
V_{\alpha {\alpha}^{'}}=\cases{
(-1)^{\frac{1-p_{\alpha}^{(l_0)}}{2}} \;\;\;\;\;\;\;\;\;\;\;\;\;\;\;\;\;\;\;\;\;\;\;\;\;\;\;\;\;\;\;\;\;\;\;\;\;\;\;\;\;\;\;\;\;\;\;\;\;\;\;\;\; \mbox{for} \;\; \alpha =1 \cr
(-1)^{\frac{1-p_{\alpha}^{(l_0)}}{2}} q^{\left( \alpha -1-p_1^{(l_0)}-p_{\alpha}^{(l_0)} -2\displaystyle \sum_{\beta =2}^{\alpha-1} p_{\beta}^{(l_0)} \right) }\;\;\;\;\;\;\;\;\;\;\;\mbox{for} \; 1<\alpha <\frac{N_0+1}{2} \cr
(-1)^{\frac{1-p_{\alpha}^{(l_0)}}{2}} q^{\left( \alpha -2-p_1^{(l_0)}-p_{\alpha}^{(l_0)} -2\displaystyle \sum_{\stackrel{\beta =2}{\neq \frac{N_0}{2}+1}}^{\alpha -1} p_{\beta}^{(l_0)} \right)  } \;\;\;\;\;\;\;\mbox{for} \; \frac{N_0+1}{2} <\alpha \leq N_0 \cr}
\end{eqnarray}
\item $U_{q}\left[ osp(2n|2m)^{(2)} \right]$:
\begin{eqnarray}
V_{\alpha {\alpha}^{'}}=\cases{
(-1)^{\frac{1-p_{\alpha}^{(l_0)}}{2}} \;\;\;\;\;\;\;\;\;\;\;\;\;\;\;\;\;\;\;\;\;\;\;\;\;\;\;\;\;\;\;\;\;\;\;\;\;\;\;\;\;\;\;\;\;\;\;\;\;\;\;\;\; \mbox{for} \;\; \alpha=1 \cr
(-1)^{\frac{1-p_{\alpha}^{(l_0)}}{2}} q^{\left( \alpha -1-p_1^{(l_0)}-p_{\alpha}^{(l_0)} -2\displaystyle \sum_{\beta=2}^{\alpha-1} p_{\beta}^{(l_0)} \right) } \;\;\;\;\;\;\;\;\;\;\;\mbox{for} \;  1<\alpha <\frac{N_0+1}{2} \cr
(-1)^{\frac{1-p_{\alpha}^{(l_0)}}{2}} q^{\left( \alpha -p_1^{(l_0)}-p_{\alpha}^{(l_0)} -2\displaystyle \sum_{\stackrel{\beta=2}{\neq \frac{N_0}{2}+1}}^{\alpha-1} p_{\beta}^{(l_0)} \right) } \;\;\;\;\;\;\;\;\;\;\;\mbox{for} \; \frac{N_0+1}{2} <\alpha \leq N_0 \cr}
\end{eqnarray}
\item $U_{q}\left[ osp(2n+1|2m)^{(1)} \right]$ and $U_{q}\left[sl(2n+1|2m)^{(2)} \right]$:
\begin{eqnarray}
V_{\alpha {\alpha}^{'}}=\cases{
(-1)^{\frac{1-p_{\alpha}^{(l_0)}}{2}}  \;\;\;\;\;\;\;\;\;\;\;\;\;\;\;\;\;\;\;\;\;\;\;\;\;\;\;\;\;\;\;\;\;\;\;\;\;\;\;\;\;\;\;\;\;\;\;\;\;\;\;\;\; \mbox{for} \;\;      \alpha=1 \cr
(-1)^{\frac{1-p_{\alpha}^{(l_0)}}{2}} q^{\left( \alpha -1-p_1^{(l_0)}-p_{\alpha}^{(l_0)} -2\displaystyle \sum_{\beta =2}^{\alpha -1} p_{\beta}^{(l_0)} \right) } \;\;\;\;\;\;\;\;\;\;\; \mbox{for} \; \; 1<\alpha <\frac{N_0+1}{2} \cr
(-1)^{\frac{1-p_{\alpha}^{(l_0)}}{2}} q^{\left( \frac{N_0}{2}-1-p_1^{(l_0)}-p_{\alpha}^{(l_0)} -2\displaystyle \sum_{\beta =2}^{\frac{N_0-1}{2}} p_{\beta}^{(l_0)} \right) } \;\;\;\;\;\;\;\; \mbox{for} \; \; \alpha =\frac{N_0+1}{2}   \cr
(-1)^{\frac{1-p_{\alpha}^{(l_0)}}{2}} q^{ \left( \alpha -2-p_1^{(l_0)}-p_{\alpha}^{(l_0)} -2\displaystyle \sum_{\beta=2}^{\alpha-1} p_{\beta}^{(l_0)} \right)  } \;\;\;\;\;\;\;\;\;\;\;  \mbox{for} \; \; \frac{N_0+1}{2} <\alpha \leq N_0 \cr }
\end{eqnarray}
\end{itemize}

\section*{\bf Appendix B : Commutations rules}
\setcounter{equation}{0}
\renewcommand{\theequation}{B.\arabic{equation}}

This appendix is devoted to complement the commutation relations presented in the main text that
are needed in the solution of the transfer matrix eigenvalue problem. The first set is
between the diagonal fields and the scalar creation
field $F(\mu)$. The relations among $B(\lambda)$ and $D(\lambda)$ with $F(\mu)$ comes directly from
the Yang-Baxter algebra but that
between $\hat{A}(\lambda)$ and $F(\mu)$ requires also the knowledge of the
commutation rule between $\vec{B}(\lambda)$ and $\vec{B}^{*} (\mu)$. They have the following form
\begin{eqnarray}
B(\mu) \stackrel{s_1}{\otimes} F(\lambda)&=& \frac{a^{(l_{0})}_{1}(\lambda -\mu)}{d^{(l_{0})}_{N_{0},N_{0}} (\lambda -\mu)} F(\lambda) \stackrel{s_1}{\otimes} B(\mu) -
\frac{d^{(l_{0})}_{1,N_{0}} (\lambda -\mu)}{d^{(l_{0})}_{N_{0},N_{0}} (\lambda -\mu)} F(\mu) \stackrel{s_1}{\otimes} B(\lambda) \nonumber \\
&-& \frac{(-1)^{p_{12}^{(l_0)}}}{d^{(l_{0})}_{N_{0},N_{0}} (\lambda -\mu)} \left[ \vec{B}(\mu) \stackrel{s_1}{\otimes} \vec{B}(\lambda) \right] \cdot [{\vec{\xi}}^{(l_{0})}_{3}(\lambda -\mu)]^{t} , \\
D(\lambda) \stackrel{s_1}{\otimes} F(\mu) &=& \frac{a^{(l_{0})}_{1}(\lambda -\mu)}{d^{(l_{0})}_{N_{0},N_{0}} (\lambda -\mu)} F(\mu) \stackrel{s_1}{\otimes} D(\lambda) - \frac{d^{(l_{0})}_{N_{0},1} (\lambda -\mu)}{d^{(l_{0})}_{N_0 ,N_0} (\lambda -\mu)} F(\lambda) \stackrel{s_1}{\otimes} D(\mu) \nonumber \\
&-&\frac{
{\vec{\xi}}^{(l_{0})}_{1}(\lambda -\mu)} 
{d^{(l_{0})}_{N_{0},N_{0}}(\lambda -\mu)} 
\cdot 
\left[ \vec{B}^{*}(\lambda) \stackrel{s_1}{\otimes} \vec{B}^{*}(\mu) \right] ,
\end{eqnarray}
\begin{eqnarray}
\hat{A}(\lambda) \stackrel{s_1}{\otimes} F(\mu) &=& \left[1-\frac{c^{(l_{0})}(\lambda -\mu) \bar{c}^{(l_{0})}(\lambda -\mu)}{(b^{(l_{0})} (\lambda -\mu))^{2}} \right] F(\mu) \stackrel{s_1}{\otimes} \hat{A}(\lambda)
+\left[ \frac{{\bar{c}}^{(l_{0})} (\lambda -\mu)}{b^{(l_{0})} (\lambda -\mu)} \right]^{2} F(\lambda) \stackrel{s_1}{\otimes} \hat{A}(\mu) \nonumber \\
&-& \frac{{\bar{c}}^{(l_{0})}(\lambda -\mu)}{b^{(l_{0})}(\lambda -\mu)}
\left[\vec{B}(\lambda) \stackrel{s_1}{\otimes} \vec{B}^{*}(\mu) - (-1)^{p_{12}^{(l_0)}} \vec{B}^{*}(\lambda) \stackrel{s_1}{\otimes} \vec{B}(\mu) \right],
\end{eqnarray}
where ${\vec{\xi}}^{(l_{0})}_{3}(\lambda)=\displaystyle
\sum_{a=1}^{N_0-2} d^{(l_{0})}_{a+1,N_0} (\lambda) \; \hat{e}_{N_0-1-a} \otimes \hat{e}_{a}$.

On the other hand the
commutation relations between the creation fields are given by
\begin{eqnarray}
{\vec B}(\lambda) \stackrel{s_1}{\otimes} {\vec B}(\mu)&=&{\vec B}(\mu) \stackrel{s_1}{\otimes} {\vec B}(\lambda) \cdot \frac{\check{r}^{(l_1)}_{12}(\lambda-\mu)}{a^{(l_{0})}_{1}(\lambda-\mu) }
+\frac{(-1)^{p_{12}^{(l_0)}}}{d^{(l_{0})}_{N_{0},N_{0}}(\lambda-\mu)} {\vec{\xi}}^{(l_{0})}_{1}(\lambda-\mu) F(\lambda) B(\mu) \nonumber \\
&+&\frac{(-1)^{p_{12}^{(l_0)}}}{a^{(l_{0})}_{1}(\lambda-\mu)} 
{\vec{\xi}}^{(l_{0})}_{2}(\lambda-\mu) F(\mu) B(\lambda) ,
\end{eqnarray}
\begin{eqnarray}
\left[ F(\lambda), F(\mu) \right] =0 ,
\end{eqnarray}
\begin{eqnarray}
F(\mu) \stackrel{s_1}{\otimes} \vec{B}(\lambda) &=& (-1)^{p_{12}^{(l_0)}}\frac{a^{(l_{0})}_{1}(\lambda-\mu)}
{b^{(l_{0})}(\lambda-\mu)} \vec{B}(\lambda) 
\stackrel{s_1}{\otimes} F(\mu) -(-1)^{p_{12}^{(l_0)}}\frac{{\bar{c}}^{(l_{0})}(\lambda-\mu)}
{b^{(l_{0})}(\lambda-\mu)} \vec{B}(\mu) \stackrel{s_1}{\otimes} F(\lambda) , \nonumber \\
\\
\vec{B}(\mu) \stackrel{s_1}{\otimes} F(\lambda) &=& (-1)^{p_{12}^{(l_0)}}\frac{a^{(l_{0})}_{1}(\lambda-\mu)}{b^{(l_{0})}(\lambda-\mu)} F(\lambda) \stackrel{s_1}{\otimes} \vec{B}(\mu) -(-1)^{p_{12}^{(l_0)}}\frac{{c}^{(l_{0})}(\lambda-\mu)}{b^{(l_{0})}(\lambda-\mu)} F(\mu) \stackrel{s_1}{\otimes} \vec{B}(\lambda). \nonumber \\
\label{comutf}
\end{eqnarray}

There are other commutation rules that either important to write appropriate relations between the
diagonal and the creation fields  or to disentangle the eigenvalue problem. They are listed below
\begin{eqnarray}
B(\lambda) \stackrel{s_1}{\otimes} \vec{B}^{*}(\mu) &=& (-1)^{p_{12}^{(l_0)}} \frac{b^{(l_0)}(\mu-\lambda)}{d_{N_0,N_0}^{(l_0)}(\mu-\lambda)}
\vec{B}^{*}(\mu) \stackrel{s_1}{\otimes} B(\lambda) 
-\vec{B}^{*} (\lambda) \stackrel{s_1}{\otimes} \hat{A} (\mu) \cdot
\frac{
[\vec{\xi}^{(l_{0})}_{3}(\mu-\lambda)]^{t}}
{d_{N_0 ,N_0}^{(l_0)}(\mu-\lambda)} 
\nonumber \\
&+& \frac{{\bar{c}}^{(l_0)}(\mu-\lambda)}{d_{N_0,N_0}^{(l_0)}(\mu-\lambda)}
F(\mu) \stackrel{s_1}{\otimes} \vec{C}(\lambda) - \frac{d_{1,N_0}^{(l_0)}(\mu-\lambda)}{d_{N_0,N_0}^{(l_0)}(\mu-\lambda)}
F(\lambda) \stackrel{s_1}{\otimes} \vec{C}(\mu)
\end{eqnarray}

\begin{eqnarray}
\vec{B}(\lambda) \stackrel{s_1}{\otimes} \vec{B}^{*}(\mu)=  (-1)^{p_{12}^{(l_0)}} \vec{B}^{*}(\mu) \stackrel{s_1}{\otimes} \vec{B}(\lambda)
+\frac{{\bar{c}}^{(l_0)}(\mu-\lambda)}{b^{(l_0)}(\mu-\lambda)} F (\mu) \stackrel{s_1}{\otimes} \hat{A} (\lambda)
- \frac{c^{(l_0)}(\mu-\lambda)}{b^{(l_0)}(\mu-\lambda)} F (\lambda) \stackrel{s_1}{\otimes} \hat{A} (\mu)
\nonumber \\
\end{eqnarray}

\begin{eqnarray}
(-1)^{p_{12}^{(l_0)}} \vec{C}(\lambda) \stackrel{s_1}{\otimes} \vec{B}(\mu) =  \vec{B}(\mu) \stackrel{s_1}{\otimes} \vec{C}(\lambda)
+\frac{{\bar{c}}^{(l_0)}(\lambda-\mu)}{b^{(l_0)}(\lambda-\mu)} \left[  B(\mu) \stackrel{s_1}{\otimes} \hat{A} (\lambda) -  B(\lambda) \stackrel{s_1}{\otimes} \hat{A} (\mu) \right]
\end{eqnarray}

\begin{eqnarray}
\vec{C}^{*} (\lambda) \stackrel{s_1}{\otimes} \vec{B}(\mu) &=& 
\vec{B}(\mu) \stackrel{s_1}{\otimes} \vec{C}^{*} (\lambda) \cdot 
\frac{\check{{\mathcal R}}_{12}(\lambda-\mu)}
{d_{N_0,N_0}^{(l_0)}(\lambda-\mu)}
- \frac{d_{N_0 ,1}^{(l_0)}(\lambda-\mu)}{d_{N_0,N_0}^{(l_0)}(\lambda-\mu)}
\vec{B}(\lambda) \stackrel{s_1}{\otimes} \vec{C}^{*} (\mu) \nonumber \\
&-& (-1)^{p_{12}^{(l_0)}} \frac{\vec{\xi}^{(l_{0})}_{4}(\lambda-\mu)}{d_{N_0,N_0}^{(l_0)}(\lambda-\mu)} \cdot
\hat{A} (\lambda) \stackrel{s_1}{\otimes} \hat{A} (\mu) + (-1)^{p_{12}^{(l_0)}} 
\frac{\vec{\xi}^{(l_{0})}_{4}(\lambda-\mu)}{d_{N_0,N_0}^{(l_0)}(\lambda-\mu)}
B(\mu) D(\lambda) \nonumber \\
&+& (-1)^{p_{12}^{(l_0)}} \frac{\vec{\xi}^{(l_{0})}_{5}(\lambda-\mu)}{d_{N_0,N_0}^{(l_0)}(\lambda-\mu)} F(\mu) C(\lambda)
\end{eqnarray}

\n where  the vectors 
$\vec{\xi}^{(l_{0})}_{4}(\lambda)$ and
$\vec{\xi}^{(l_{0})}_{5}(\lambda)$ are

\begin{equation}
\vec{\xi}^{(l_{0})}_{4} (\lambda) = \sum_{a=1}^{N_{0} -2} d^{(l_0)}_{N_0 , N_0 -a} (\lambda) \hat{e}_{N_0 -1 -a} \otimes \hat{e}_{a}
\end{equation}

\begin{equation}
\vec{\xi}^{(l_{0})}_{5} (\lambda) = \sum_{a=1}^{N_{0} -2} d^{(l_0)}_{1 , N_0 -a} (\lambda) \hat{e}_{N_0 -1 -a} \otimes \hat{e}_{a}
\end{equation}

Finally, the  matrix $\check{{\mathcal R}}_{12}(\lambda) $  can be represented by 
\begin{equation}
\check{{\mathcal R}}_{12}(\lambda)=\sum_{abcd}^{N_0-2} R^{a+1,c+1}_{b+1,d+1} (\lambda) \hat{e}_{ab}^{(1)} \otimes \hat{e}_{cd}^{(2)}
\end{equation}
where  $R^{a,c}_{b,d}$ are the matrix elements of $\check{R}^{(l_0)}_{12}(\lambda)$. Here we recall that
we have used the convention
$\displaystyle \check{R}^{(l_0)}_{12}(\lambda)=\sum_{abcd}^{N_0} R^{a,c}_{b,d} 
(\lambda) \hat{e}_{ab}^{(1)} \otimes \hat{e}_{cd}^{(2)}$.

\section*{\bf Appendix C : Auxiliary Bethe Ansatz}
\setcounter{equation}{0}
\renewcommand{\theequation}{C.\arabic{equation}}

In this appendix we present the Bethe ansatz results concerning the last step $l_f$ of the nested
construction  presented in section 4.  In general, we need to diagonalize the following inhomogeneous
transfer matrix
\begin{eqnarray}
&& T^{(l_f)}(\lambda,\{\lambda_1^{(l_f)}, \dots, \lambda_{m_{l_{f}}}^{(l_f)} \}) \nonumber \\
&& =Str_{{\mathcal A}^{(f)}} \left[ R^{(l_f)}_{{\mathcal A}^{(f)} m_{l_f}} (\lambda-\lambda^{(l_f)}_{m_{l_f}})  R^{(l_f)}_{{\mathcal A}^{(f)} m_{l_f}-1} (\lambda-\lambda^{(l_f)}_{m_{l_f}-1})
\dots R^{(l_f)}_{{\mathcal A}^{(f)} 1} (\lambda-\lambda^{(l_f)}_{1}) \right]
\end{eqnarray}

We now begin to list the respective $R$-matrix
$R^{(l_f)}_{{\mathcal A}^{(f)} j} (\lambda)$, the eigenvalue expression \\
$\Lambda^{(l_f)}(\lambda,\{\lambda_1^{(l_f)}, \dots, \lambda_{m_{l_{f}}}^{(l_f)} \})$ and the corresponding Bethe ansatz equations of the vertex models mentioned in Table 2.
The
$U_q[sl(2n+1|2m)^{(2)}]$ and $U_q[osp(2n+1|2m)^{(1)}]$  vertex models
have $l_{f}=(3|0)$ and the corresponding $R^{(l_f)} (\lambda)$ is given by

\begin{eqnarray}
&& R^{(l_{f})}(\lambda) \nonumber \\
&& =\pmatrix{
a_{1}^{(l_{f})}(\lambda) & 0 & 0 & 0 & 0 & 0 & 0 & 0 & 0 \cr
0 & b^{(l_{f})}(\lambda) & 0 & c^{(l_{f})}(\lambda) & 0 & 0 & 0 & 0 & 0 \cr
0 & 0 & d_{1,1}^{(l_{f})}(\lambda) & 0 & d_{1,2}^{(l_{f})}(\lambda) & 0 & d_{1,3}^{(l_{f})}(\lambda) & 0 & 0 \cr
0 & {\bar{c}}^{(l_{f})}(\lambda) & 0 & b^{(l_{f})}(\lambda) & 0 & 0 & 0 & 0 & 0 \cr
0 & 0 & d_{2,1}^{(l_{f})}(\lambda) & 0 & d_{2,2}^{(l_{f})}(\lambda) & 0 & d_{2,3}^{(l_{f})}(\lambda) & 0 & 0 \cr
0 & 0 & 0 & 0 & 0 & b^{(l_{f})}(\lambda) & 0 & c^{(l_{f})}(\lambda) & 0 \cr
0 & 0 & d_{3,1}^{(l_{f})}(\lambda) & 0 & d_{3,2}^{(l_{f})}(\lambda) & 0 & d_{3,3}^{(l_{f})}(\lambda) & 0 & 0 \cr
0 & 0 & 0 & 0 & 0 & {\bar{c}}^{(l_{f})}(\lambda) & 0 & b^{(l_{f})}(\lambda) & 0 \cr
0 & 0 & 0 & 0 & 0 & 0 & 0 & 0 & a_{1}^{(l_{f})}(\lambda) \cr} \nonumber \\
\end{eqnarray}

\n where the Boltzmann weights for each superalgebra are given by the set of relations (\ref{bw1} - \ref{bwf}).

For the $U_q[sl(2n+1|2m)^{(2)}]$ vertex model the underlying $R^{(l_f)} (\lambda)$ operator is related to that
of the Izergin-Korepin \cite{IK} model and has the following expressions for the 
eigenvalues and Bethe Ansatz equations

\begin{eqnarray}
\Lambda^{(l_{f})} (\lambda , \{ \lambda_{1}^{(l_{f})},\dots,\lambda_{m_{l_{f}}}^{(l_{f})} \} ) &=& \prod_{i=1}^{m_{l_{f}}} a_{1}^{(l_{f})}(\lambda -\lambda_{i}^{(l_{f})} )
\frac{Q_{f+1}\left( \lambda + i\gamma \right)}{Q_{f+1}\left( \lambda  \right)} \nonumber \\
&+& \prod_{i=1}^{m_{l_{f}}} d_{N_{f},N_{f}}^{(l_{f})}(\lambda -\lambda_{i}^{(l_{f})} )
\frac{Q_{f+1}\left( \lambda -i \frac{3\gamma}{2}  +i\frac{\pi}{2} \right)}{Q_{f+1}\left( \lambda - i\frac{\gamma}{2} +i\frac{\pi}{2} \right)} \nonumber \\
&+& \prod_{i=1}^{m_{l_{f}}} b^{(l_{f})} (\lambda -\lambda_{i}^{(l_{f})} ) \frac{Q_{f+1}\left( \lambda - i\gamma \right)}{Q_{f+1}\left( \lambda  \right)} \frac{Q_{f+1}\left( \lambda +i \frac{\gamma}{2} +i\frac{\pi}{2} \right)}{Q_{f+1}\left( \lambda - i\frac{\gamma}{2} +i\frac{\pi}{2} \right)} ,
\end{eqnarray}

\begin{eqnarray}
\prod_{i=1}^{m_{l_{f}}} \frac{\sinh{\left(\lambda_{j}^{(l_{f+1})} -\lambda_{i}^{(l_{f})} -i \gamma \right)}}{\sinh{\left(\lambda_{j}^{(l_{f+1})} -\lambda_{i}^{(l_{f})} \right)}}&=&
\prod_{i\neq j}^{m_{l_{f+1}}} \frac{\sinh{\left(\lambda_{j}^{(l_{f+1})} -\lambda_{i}^{(l_{f+1})} -i\gamma \right)}}{\sinh{\left(\lambda_{j}^{(l_{f+1})} -\lambda_{i}^{(l_{f+1})} +i\gamma \right)}}
\frac{\cosh{\left(\lambda_{j}^{(l_{f+1})} -\lambda_{i}^{(l_{f+1})} +i\frac{\gamma}{2} \right)}}{\cosh{\left(\lambda_{j}^{(l_{f+1})} -\lambda_{i}^{(l_{f+1})} - i\frac{\gamma}{2} \right)}} . \nonumber \\
\end{eqnarray}

For the $U_q[osp(2n+1|2m)^{(1)}]$ vertex model, however, the 
underlying $R^{(l_f)} (\lambda)$ operator is similar
to that
of the Fateev-Zamolodchikov \cite{FZ} model and 
expressions for the eigenvalues and Bethe Ansatz equations are given by

\begin{eqnarray}
\Lambda^{(l_{f})} (\lambda | \{ \lambda_{1}^{(l_{f})},\dots,\lambda_{m_{l_{f}}}^{(l_{f})} \} ) &=& \prod_{i=1}^{m_{l_{f}}} a_{1}^{(l_{f})}(\lambda -\lambda_{i}^{(l_{f})} )
\frac{Q_{f+1}\left( \lambda + i\gamma \right)}{Q_{f+1}\left( \lambda  \right)} \nonumber \\
&+& \prod_{i=1}^{m_{l_{f}}} d_{N_{f},N_{f}}^{(l_{f})}(\lambda -\lambda_{i}^{(l_{f})} )
\frac{Q_{f+1}\left( \lambda - i\frac{\gamma}{2} \right)}{Q_{f+1}\left( \lambda + i\frac{\gamma}{2} \right)} \nonumber \\
&+& \prod_{i=1}^{m_{l_{f}}} b^{(l_{f})} (\lambda -\lambda_{i}^{(l_{f})} ) \frac{Q_{f+1}\left( \lambda + i\gamma \right)}{Q_{f+1}\left( \lambda  \right)} \frac{Q_{f+1}\left( \lambda - i\frac{\gamma}{2} \right)}{Q_{f+1}\left( \lambda + i\frac{\gamma}{2} \right)} ,
\end{eqnarray}

\begin{eqnarray}
\prod_{i=1}^{m_{l_{f}}} \frac{\sinh{\left(\lambda_{j}^{(l_{f+1})} -\lambda_{i}^{(l_{f})} - i\gamma \right)}}{\sinh{\left(\lambda_{j}^{(l_{f+1})} -\lambda_{i}^{(l_{f})} \right)}}&=&
\prod_{i\neq j}^{m_{l_{f+1}}} \frac{\sinh{\left(\lambda_{j}^{(l_{f+1})} -\lambda_{i}^{(l_{f+1})} -i\frac{\gamma}{2} \right)}}{\sinh{\left(\lambda_{j}^{(l_{f+1})} -\lambda_{i}^{(l_{f+1})} +i\frac{\gamma}{2} \right)}} .
\end{eqnarray}

The last step for the $U_q[sl(2n|2m)^{(2)}]$ and $U_q[osp(2n|2m)^{(2)}]$ is $l_{f} = (2|0)$
and the form of the underlying R-matrix is that of the six-vertex model, namely
\begin{equation}
R^{(l_{f})}(\lambda)=\pmatrix{
a_{1}^{(l_{f})}(\lambda) & 0 & 0 & 0  \cr
0 & d_{1,1}^{(l_{f})}(\lambda) & d_{1,2}^{(l_{f})}(\lambda) & 0  \cr
0 & d_{2,1}^{(l_{f})}(\lambda) & d_{2,2}^{(l_{f})}(\lambda) & 0  \cr
0 & 0 & 0 & a_{1}^{(l_{f})}(\lambda) \cr}
\end{equation}

\n where each superalgebra has its own Boltzmann weights defined in (\ref{bw1} - \ref{bwf}).

It turns out that for the $U_q[sl(2n|2m)^{(2)}]$ vertex models the last step has the following eigenvalues 
and Bethe Ansatz
equations,

\begin{eqnarray}
\Lambda^{(l_{f})} (\lambda , \{ \lambda_{1}^{(l_{f})},\dots,\lambda_{m_{l_{f}}}^{(l_{f})} \} ) &=& \prod_{i=1}^{m_{l_{f}}} a_{1}^{(l_{f})}(\lambda -\lambda_{i}^{(l_{f})} )
\frac{Q_{f+1}\left( \lambda + i\gamma \right)}{Q_{f+1}\left( \lambda  \right)} \frac{Q_{f+1}\left( \lambda +i \gamma +i\frac{\pi}{2}   \right)}{Q_{f+1}\left( \lambda +i\frac{\pi}{2} \right)}\nonumber \\
&+& \prod_{i=1}^{m_{l_{f}}} d_{N_{f},N_{f}}^{(l_{f})}(\lambda -\lambda_{i}^{(l_{f})} )
\frac{Q_{f+1}\left( \lambda -  i\gamma  \right)}{Q_{f+1}\left( \lambda  \right)} \frac{Q_{f+1}\left( \lambda -  i\gamma +i\frac{\pi}{2} \right)}{Q_{f+1}\left( \lambda +i\frac{\pi}{2} \right)} ,
\end{eqnarray}

\begin{eqnarray}
\prod_{i=1}^{m_{l_{f}}} \frac{\sinh{\left[2 \left(   \lambda_{j}^{(l_{f+1})} -\lambda_{i}^{(l_{f})} -  i\gamma \right) \right]}}{\sinh{\left[2 \left( \lambda_{j}^{(l_{f+1})} -\lambda_{i}^{(l_{f})} \right) \right]}}&=&
\prod_{i\neq j}^{m_{l_{f+1}}} \frac{\sinh{\left[ 2 \left( \lambda_{j}^{(l_{f+1})} -\lambda_{i}^{(l_{f+1})} -i\gamma \right) \right]}}{\sinh{\left[2 \left( \lambda_{j}^{(l_{f+1})} -\lambda_{i}^{(l_{f+1})} +i\gamma \right) \right]}} .
\end{eqnarray}
while for the $U_q[osp(2n|2m)^{(2)}]$ vertex models the last step results are 

\begin{eqnarray}
\Lambda^{(l_{f})} (\lambda , \{ \lambda_{1}^{(l_{f})},\dots,\lambda_{m_{l_{f}}}^{(l_{f})} \} ) &=& \prod_{i=1}^{m_{l_{f}}} a_{1}^{(l_{f})}(\lambda -\lambda_{i}^{(l_{f})} )
\frac{Q_{f+1}\left( \lambda + 2i\gamma \right)}{Q_{f+1}\left( \lambda  \right)} \nonumber \\
&+& \prod_{i=1}^{m_{l_{f}}} d_{N_{f},N_{f}}^{(l_{f})}(\lambda -\lambda_{i}^{(l_{f})} )
\frac{Q_{f+1}\left( \lambda - 2i \gamma  \right)}{Q_{f+1}\left( \lambda  \right)} ,
\end{eqnarray}

\begin{eqnarray}
\prod_{i=1}^{m_{l_{f}}} \frac{\sinh{\left(\lambda_{j}^{(l_{f+1})} -\lambda_{i}^{(l_{f})} - 2i \gamma \right)}}{\sinh{\left(\lambda_{j}^{(l_{f+1})} -\lambda_{i}^{(l_{f})} \right)}}&=&
\prod_{i\neq j}^{m_{l_{f+1}}} \frac{\sinh{\left(\lambda_{j}^{(l_{f+1})} -\lambda_{i}^{(l_{f+1})} -2i\gamma \right)}}{\sinh{\left(\lambda_{j}^{(l_{f+1})} -\lambda_{i}^{(l_{f+1})} +2i\gamma \right)}} .
\end{eqnarray}

For the $U_q[osp(2n|2m)^{(1)}]$ vertex models the last step occurs at $l_{f} = (4|0)$ and a more
careful analysis is required. First
we perform the 
transformation $R^{(l_{f})}_{{\mathcal A}^{(f)} j} \rightarrow M_{j}^{-1} R^{(l_{f})}_{{\mathcal A}^{(f)} j}  M_{j}$,
that preserves the spectrum of the transfer matrix associated. For each $j$-th site of the lattice the
matrix $M_j$ is
\begin{equation}
M_{j}=\pmatrix{
1 & 0 & 0 & 0 \cr
0 & 1 & 0 & 0 \cr
0 & 0 & 1 & 0 \cr
0 & 0 & 0 & -1 \cr}_{j}
\end{equation}

Now it is not difficult to show that 
the transformed $R^{(l_{f})}_{{\mathcal A}^{(f)} j}$ matrix can be decomposed in terms of the tensor product of
two 6-vertex models. More precisely, the new $R^{(l_{f})}$ can be written as
\begin{equation}
R^{(l_{f})}(\lambda)= R^{6v}_{\sigma} (\lambda) \; R^{6v}_{\tau} (\lambda)
\end{equation}

\n where $\sigma$ and $\tau$ represent two commuting basis. These basis can be easily constructed in terms of
Pauli matrices by the following relations
\begin{eqnarray}
\sigma^{\alpha} = \sigma^{\alpha}_{P} \otimes I_{2\times 2} , \nonumber \\
\\
\tau^{\alpha} = I_{2\times 2} \otimes \sigma^{\alpha}_{P} , \nonumber
\end{eqnarray}

\n where $\sigma^{\alpha}$ and $\tau^{\alpha}$ are elements of $\sigma$ and $\tau$ basis respectively,
$\sigma^{\alpha}_{P}$ is a Pauli matrix (with the identity included) and $I_{2\times 2}$ is the
identity matrix with dimensions $2\times 2$.

\n Performing the above procedure we have the following expressions for the R-matrix of these two $6$-vertex models,

\begin{equation}
\label{6va}
R^{6v}_{\sigma}(\lambda)=\pmatrix{
a_{\sigma}(\lambda) & 0 & 0 & 0  \cr
0 & b_{\sigma}(\lambda) & \bar{c}_{\sigma}(\lambda) & 0  \cr
0 & c_{\sigma}(\lambda) & b_{\sigma}(\lambda) & 0  \cr
0 & 0 & 0 & a_{\sigma}(\lambda) \cr} ,
\;\;\;\;\;\;\;
R^{6v}_{\tau}(\lambda)=\pmatrix{
a_{\tau}(\lambda) & 0 & 0 & 0  \cr
0 & b_{\tau}(\lambda) & \bar{c}_{\tau}(\lambda) & 0  \cr
0 & c_{\tau}(\lambda) & b_{\tau}(\lambda) & 0  \cr
0 & 0 & 0 & a_{\tau}(\lambda) \cr} ,
\end{equation}

\n whose Boltzmann weights are given by

\begin{eqnarray}
\label{6vb}
a_{\sigma}(\lambda)&=&1 \;\;\;\;\;\;\;\;\;\;\;\;\;\;\;\;\;\;\;\;\;\;\;\;\;\;\; a_{\tau}(\lambda)=(e^{2 \lambda}-q^2)^{2} \nonumber \\
b_{\sigma}(\lambda)&=&q\frac{(e^{2 \lambda}-1)}{(e^{2 \lambda}-q^2)} \;\;\;\;\;\;\;\;\;\;\; b_{\tau}(\lambda)=q(e^{2 \lambda}-1)(e^{2 \lambda}-q^2) \nonumber \\
c_{\sigma}(\lambda)&=&\frac{(1-q^{2})}{(e^{2 \lambda}-q^2)} \;\;\;\;\;\;\;\;\;\;\;\;\; c_{\tau}(\lambda)=e^{2 \lambda} (1-q^{2})(e^{2 \lambda}-q^2)\nonumber \\
\bar{c}_{\sigma}(\lambda)&=&e^{2\lambda} \frac{(1-q^{2})}{(e^{2 \lambda}-q^2)} \;\;\;\;\;\;\;\;\; \bar{c}_{\tau}(\lambda)=(1-q^{2})(e^{2 \lambda}-q^2) . \nonumber \\
\end{eqnarray}

Consequently, the transfer matrix eigenvalues associated to $R^{(l_{f})}$ vertex model can
be decomposed as a product of the eigenvalues of two 6-vertex models  defined in Eq.(\ref{6va}-\ref{6vb}).
In the presence of inhomogeneities 
$\{ \lambda_{1}^{(l_{f})},\dots, \lambda_{m_{l_{f}}}^{(l_{f})} \}$ we find that these eigenvalues are
given by

\begin{eqnarray}
\Lambda^{(l_{f})} (\lambda , \{ \lambda_{1}^{(l_{f})},\dots,\lambda_{m_{l_{f}}}^{(l_{f})} \} ) &=& \left[ \frac{Q_{+}\left( \lambda + i\gamma \right)}{Q_{+}\left( \lambda  \right)}
+ \prod_{i=1}^{m_{l_{f}}} b_{\sigma}(\lambda -\lambda_{i}^{(l_{f})} )
\frac{Q_{+}\left( \lambda - i\gamma  \right)}{Q_{+}\left( \lambda  \right)} \right] \nonumber \\
&\times& \left[ \prod_{i=1}^{m_{l_{f}}} a_{\tau}(\lambda -\lambda_{i}^{(l_{f})} )\frac{Q_{-}\left( \lambda + i\gamma \right)}{Q_{-}\left( \lambda  \right)}
+ \prod_{i=1}^{m_{l_{f}}} b_{\tau}(\lambda -\lambda_{i}^{(l_{f})} )
\frac{Q_{-}\left( \lambda - i\gamma  \right)}{Q_{-}\left( \lambda  \right)} \right] , \nonumber \\
\end{eqnarray}

\n and the Bethe Ansatz equations are

\begin{eqnarray}
\prod_{i=1}^{m_{l_{f}}} \frac{\sinh{\left(\lambda_{j}^{(l_{\pm})} -\lambda_{i}^{(l_{f})} - i\gamma \right)}}{\sinh{\left(\lambda_{j}^{(l_{\pm})} -\lambda_{i}^{(l_{f})} \right)}}=
\prod_{i\neq j}^{m_{l_{\pm}}} \frac{\sinh{\left(\lambda_{j}^{(l_{\pm})} -\lambda_{i}^{(l_{\pm})} - i\gamma \right)}}{\sinh{\left(\lambda_{j}^{(l_{\pm})} -\lambda_{i}^{(l_{\pm})} + i\gamma \right)}} .
\end{eqnarray}

\section*{\bf Appendix D : The $U_q[sl(1|2m)^{(2)}]$ and  $U_q[osp(1|2m)^{(1)}]$ results}
\setcounter{equation}{0}
\renewcommand{\theequation}{D.\arabic{equation}}

In this appendix we list  
the Bethe ansatz results for the $U_{q}[sl(1|2m)^{(2)}]$ and $U_{q}[osp(1|2m)^{(1)}]$
vertex models. In both cases the  
last step occurs at
$l_{f}=(1|2)$ and the corresponding  
$R^{(l_{f})}$ can be related to that of the Fateev-Zamolodchikov \cite{FZ} or Izergin-Korepin \cite{IK} 
vertex models, respectively. Since the problem of diagonalizing these systems have already been discussed
in the previous appendix we restrict ourselves here in presenting only the main results.
The eigenvalue expression for the $U_{q}[sl(1|2m)^{(2)}]$ vertex model is
\begin{eqnarray}
\Lambda^{(l_{0})} (\lambda) &=& - \left[ - a_{1}^{(l_{0})}(\lambda) \right]^{L}
\frac{Q_{1}\left( \lambda - i\frac{\gamma}{2} \right)}{Q_{1}\left( \lambda + i\frac{\gamma}{2} \right)}
-  \left[ - d_{N_{0},N_{0}}^{(l_{0})}(\lambda) \right]^{L}
\frac{Q_{1}\left( \lambda + i m \gamma  +i\frac{\pi}{2} \right)}{Q_{1}\left( \lambda + i\left(m-1 \right)\gamma +i\frac{\pi}{2} \right)} \nonumber \\
&+& \left[ b^{(l_{0})} (\lambda)  \right]^{L} \sum_{\alpha=1}^{N_{0} -2} G_{\alpha}(\lambda | \{ \lambda_{j}^{(l_{\beta})} \}) \nonumber \\
\end{eqnarray}
\begin{eqnarray}
&& G_{\alpha}(\lambda | \{ \lambda_{j}^{(l_{\beta})} \}) \nonumber \\
&& =\cases{
\frac{Q_{\alpha} \left(\lambda +i\left(\frac{\alpha +2}{2} \right)\gamma  \right)}{Q_{\alpha} \left(\lambda +i\frac{\alpha}{2} \gamma  \right)}
\frac{Q_{\alpha+1} \left(\lambda +i\left(\frac{\alpha -1}{2} \right)\gamma  \right)}{Q_{\alpha+1} \left(\lambda +i\left(\frac{\alpha +1}{2} \right)\gamma  \right)}
\;\;\;\;\;\;\;\;\;\;\;\;\;\;\;\;\;\;\;\;\;\;\;\;\;\;\;\;\;\;\;\;\;\;\;\;\; \alpha = 1,\dots ,m-1 \cr
\frac{Q_{\alpha} \left(\lambda +i\left(\frac{m -2}{2} \right)\gamma  \right)}{Q_{\alpha} \left(\lambda +i\frac{m}{2} \gamma \right)}
\frac{Q_{\alpha} \left(\lambda +i\left(\frac{m +1}{2} \right)\gamma +i\frac{\pi}{2} \right)}{Q_{\alpha} \left(\lambda+i \left(\frac{m-1}{2} \right)\gamma +i\frac{\pi}{2} \right)}
\;\;\;\;\;\;\;\;\;\;\;\;\;\;\;\;\;\;\;\;\;\;\;\;\;\;\;\;\;\;\;\;\;\; \alpha = m \cr
G_{\alpha - m}(-i\frac{\pi}{2} - i(m-\frac{1}{2})\gamma -\lambda |- \{ \lambda_{j}^{(l_{\beta})} \}) \;\;\;\;\;\;\;\;\;\;\;\;\;\;\;\;\;\;\;\;\;\; \alpha = m+1,\dots ,2m-1 \cr } \nonumber
\end{eqnarray}

\n and the  respective Bethe Ansatz equations are

\begin{eqnarray}
\prod_{i=1}^{m_{l_{\alpha -1}}} \frac{\sinh{\left(\lambda_{j}^{(l_{\alpha})} -\lambda_{i}^{(l_{\alpha-1})} +i\frac{\gamma}{2} \right)}}{\sinh{\left(\lambda_{j}^{(l_{\alpha})} -\lambda_{i}^{(l_{\alpha-1})} -i\frac{\gamma}{2} \right)}}&=&
\prod_{i\neq j}^{m_{l_{\alpha}}} \frac{\sinh{\left(\lambda_{j}^{(l_{\alpha})} -\lambda_{i}^{(l_{\alpha})} +i\gamma \right)}}{\sinh{\left(\lambda_{j}^{(l_{\alpha})} -\lambda_{i}^{(l_{\alpha})} -i\gamma \right)}}
\prod_{i=1}^{m_{l_{\alpha +1}}} \frac{\sinh{\left(\lambda_{i}^{(l_{\alpha+1})} -\lambda_{j}^{(l_{\alpha})} + i\frac{\gamma}{2} \right)}}{\sinh{\left(\lambda_{i}^{(l_{\alpha+1})} -\lambda_{j}^{(l_{\alpha})} - i \frac{\gamma}{2} \right)}} \nonumber \\
&&  \;\;\;\;\;\;\;\;\;\;\;\;\;\;\;\;\;\;\;\;\;\;\;\;\;\;\;\;\;\;\;\;\;\;\;\;\;\;\;\;\;\;\;\;\;\;\;\;\;\;\;\;\;\;\;\;\;\;\;\;\;\;\;\; \alpha= 1,\dots , m-1  \nonumber \\
\prod_{i=1}^{m_{l_{\alpha -1}}} \frac{\sinh{\left(\lambda_{j}^{(l_{\alpha})} -\lambda_{i}^{(l_{\alpha-1})} +i\frac{\gamma}{2} \right)}}{\sinh{\left(\lambda_{j}^{(l_{\alpha})} -\lambda_{i}^{(l_{\alpha-1})} -i\frac{\gamma}{2} \right)}}&=&
\prod_{i\neq j}^{m_{l_{\alpha}}} \frac{\cosh{\left(\lambda_{j}^{(l_{\alpha})} -\lambda_{i}^{(l_{\alpha})} +i\frac{\gamma}{2} \right)}}{\cosh{\left(\lambda_{j}^{(l_{\alpha})} -\lambda_{i}^{(l_{\alpha})} - i\frac{\gamma}{2} \right)}}
\;\;\;\;\;\;\;\;\;\;\;\;\;\;\;\;\;\;\;\;\;\;\;\;\;\;\;\;\;\;\;\;\;\;\;\;\;\; \alpha=m \nonumber \\
\end{eqnarray}

On the other hand, for the $U_{q}[osp(1|2m)^{(1)}]$ we have
\begin{eqnarray}
\Lambda^{(l_{0})} (\lambda) &=& - \left[ - a_{1}^{(l_{0})}(\lambda) \right]^{L}
\frac{Q_{1}\left( \lambda - i\frac{\gamma}{2} \right)}{Q_{1}\left( \lambda + i\frac{\gamma}{2} \right)}
-  \left[ - d_{N_{0},N_{0}}^{(l_{0})}(\lambda) \right]^{L}
\frac{Q_{1}\left( \lambda + i\left(m +1\right)\gamma  \right)}{Q_{1}\left( \lambda + i m \gamma  \right)} \nonumber \\
&+& \left[ b^{(l_{0})} (\lambda)  \right]^{L} \sum_{\alpha=1}^{N_{0} -2} G_{\alpha}(\lambda | \{ \lambda_{j}^{(l_{\beta})} \}) \nonumber \\
\end{eqnarray}

\begin{eqnarray}
&& G_{\alpha}(\lambda | \{ \lambda_{j}^{(l_{\beta})} \}) \nonumber \\
&& =\cases{
\frac{Q_{\alpha} \left(\lambda +i\left(\frac{\alpha +2}{2} \right)\gamma  \right)}{Q_{\alpha} \left(\lambda +i\frac{\alpha}{2} \gamma  \right)}
\frac{Q_{\alpha+1} \left(\lambda +i\left(\frac{\alpha -1}{2} \right)\gamma  \right)}{Q_{\alpha+1} \left(\lambda +i\left(\frac{\alpha +1}{2} \right)\gamma  \right)}
\;\;\;\;\;\;\;\;\;\;\;\;\;\;\;\;\;\;\;\;\;\;\;\;\;\;\;\;\;\;\;\;\;\;\;\;\;\;\; \alpha = 1,\dots ,m-1 \cr
\frac{Q_{\alpha} \left(\lambda +i\left(\frac{m -1}{2} \right)\gamma  \right)}{Q_{\alpha} \left(\lambda +i\left(\frac{m+1}{2} \right) \gamma \right)}
\frac{Q_{\alpha} \left(\lambda +i\left(\frac{m +2}{2} \right)\gamma  \right)}{Q_{\alpha} \left(\lambda +i\frac{m}{2} \gamma  \right)}
\;\;\;\;\;\;\;\;\;\;\;\;\;\;\;\;\;\;\;\;\;\;\;\;\;\;\;\;\;\;\;\;\;\;\;\;\;\;\;\;\; \alpha = m \cr
G_{\alpha - m}(- i(m+\frac{1}{2})\gamma -\lambda |- \{ \lambda_{j}^{(l_{\beta})} \}) \;\;\;\;\;\;\;\;\;\;\;\;\;\;\;\;\;\;\;\;\;\;\;\;\;\;\;\;\;\;\;\; \alpha = m+1,\dots ,2m-1 \cr } \nonumber
\end{eqnarray}

\n while the Bethe Ansatz equations are given by

\begin{eqnarray}
\prod_{i=1}^{m_{l_{\alpha -1}}} \frac{\sinh{\left(\lambda_{j}^{(l_{\alpha})} -\lambda_{i}^{(l_{\alpha-1})} +i\frac{\gamma}{2} \right)}}{\sinh{\left(\lambda_{j}^{(l_{\alpha})} -\lambda_{i}^{(l_{\alpha-1})} -i\frac{\gamma}{2} \right)}}&=&
\prod_{i\neq j}^{m_{l_{\alpha}}} \frac{\sinh{\left(\lambda_{j}^{(l_{\alpha})} -\lambda_{i}^{(l_{\alpha})} +i\gamma \right)}}{\sinh{\left(\lambda_{j}^{(l_{\alpha})} -\lambda_{i}^{(l_{\alpha})} -i\gamma \right)}}
\prod_{i=1}^{m_{l_{\alpha +1}}} \frac{\sinh{\left(\lambda_{i}^{(l_{\alpha+1})} -\lambda_{j}^{(l_{\alpha})} + i\frac{\gamma}{2} \right)}}{\sinh{\left(\lambda_{i}^{(l_{\alpha+1})} -\lambda_{j}^{(l_{\alpha})} - i \frac{\gamma}{2} \right)}} \nonumber \\
&& \;\;\;\;\;\;\;\;\;\;\;\;\;\;\;\;\;\;\;\;\;\;\;\;\;\;\;\;\;\;\;\;\;\;\;\;\;\;\;\;\;\;\;\;\;\;\;\;\;\;\;\;\;\;\;\;\;\;\;\;\;\;\;\; \alpha= 1,\dots , m-1  \nonumber \\
\prod_{i=1}^{m_{l_{\alpha -1}}} \frac{\sinh{\left(\lambda_{j}^{(l_{\alpha})} -\lambda_{i}^{(l_{\alpha-1})} +i\frac{\gamma}{2} \right)}}{\sinh{\left(\lambda_{j}^{(l_{\alpha})} -\lambda_{i}^{(l_{\alpha-1})} -i\frac{\gamma}{2} \right)}}&=&
\prod_{i\neq j}^{m_{l_{\alpha}}} \frac{\sinh{\left(\lambda_{j}^{(l_{\alpha})} -\lambda_{i}^{(l_{\alpha})} +i\gamma \right)}}{\sinh{\left(\lambda_{j}^{(l_{\alpha})} -\lambda_{i}^{(l_{\alpha})} -i\gamma \right)}}
\frac{\sinh{\left(\lambda_{j}^{(l_{\alpha})} -\lambda_{i}^{(l_{\alpha})} -i\frac{\gamma}{2} \right)}}{\sinh{\left(\lambda_{j}^{(l_{\alpha})} -\lambda_{i}^{(l_{\alpha})} + i\frac{\gamma}{2} \right)}}  \nonumber \\
&& \;\;\;\;\;\;\;\;\;\;\;\;\;\;\;\;\;\;\;\;\;\;\;\;\;\;\;\;\;\;\;\;\;\;\;\;\;\;\;\;\;\;\;\;\;\;\;\;\;\;\;\;\;\;\;\;\;\;\;\;\;\;\;\;\;\;\;\;\;\;\;\;\;\;\;\;\;\;\; \alpha=m \nonumber \\
\end{eqnarray}

\newpage
     
\Large
Tables 
\normalsize
\vspace{1.5cm}

\underline{Table 1}:  The values of the dimension $N_0$ and the parameter $\zeta^{(l_{0})}$. In general the
integers $n,m \geq 1$ 
except for the $U_q[osp(2n|2m)^{(1)}]$ where $n \geq 2$.

\btb[h]
\bc
\bt{|c|c|c|c|} \hline
     $U_q[{\cal{G}}]$  &$N_{0}$  &$\zeta^{(l_0)}$   \\ \hline\hline
$U_q[sl(2n+1|2m)^{(2)}]$   &$2n+2m+1$  &-$q^{2n-2m+1}$     \\ \hline
$U_q[sl(2n|2m)^{(2)}]$  &$2n+2m$  &-$q^{2n-2m}$       \\ \hline
$U_q[osp(2n|2m)^{(1)}]$  &$2n+2m$  &$q^{2n-2m-2}$    \\ \hline
$U_q[osp(2n+1|2m)^{(1)}]$  &$2n+2m+1$  &$q^{2n-2m-1}$    \\ \hline
$U_q[osp(2n|2m)^{(2)}]$  &$2n+2m$  &$q^{2n-2m+2}$    \\ \hline
\et
\ec
\etb
\vspace{1.5cm}

\underline{Table 2}: Parameters of the vertex models associated with the last step
Bethe ansatz analysis for the
$q$-deformed  Lie superalgebras. The symbols IK and FZ
stand for  Izergin-Korepin \cite{IK} and
Fateev-Zamolodchikov models \cite{FZ},
respectively.

\begin{table}
\begin{center}
\begin{tabular}{|c|c|c|} \hline
  Superalgebra     & $l_{f}$ & $R^{(l_f)}$ matrix   \\ \hline\hline
$U_q[sl(2n+1|2m)^{(2)}]$ & $(3|0)$ & nineteen-vertex IK model  \\ \hline
$U_q[sl(2n|2m)^{(2)}]$ and $U_q[osp(2n|2m)^{(2)}]$ & $(2|0)$ & six-vertex model \\ \hline
$U_q[osp(2n+1|2m)^{(1)}]$ & $(3|0)$ & nineteen-vertex FZ model  \\ \hline
$U_q[osp(2n|2m)^{(1)}]$ & $(4|0)$ & two decoupled six-vertex models  \\ \hline
\end{tabular}
\end{center}
\end{table}
\vspace{1.5cm}

\underline{Table 3}: Table with the shifts performed in the rapidities.

\begin{table}[h]
\begin{center}
\begin{tabular}{|c|c|} \hline
Lie Superalgebra & $\delta^{(l_{\alpha})}$ \\ \hline\hline
$sl(2n+1|2m)^{(2)}$, $osp(2n+1|2m)^{(1)}$  and $sl(2n|2m)^{(2)}$ & $\cases{
i\frac{\alpha}{2}\gamma \;\;\;\;\;\;\;\;\;\;\;\;\;\;\;\;\;\;\;\; 1 \leq \alpha \leq m \cr
i\left(m -\frac{\alpha}{2}\right) \gamma \;\;\;\;\;\;\;\; m < \alpha \leq m+n \cr}$ \\ \hline
$osp(2n|2m)^{(2)}$ & $\cases{
i\frac{\alpha}{2}\gamma \;\;\;\;\;\;\;\;\;\;\;\;\;\;\;\;\;\;\;\; 1 \leq \alpha \leq m \cr
i\left(m -\frac{\alpha}{2}\right) \gamma \;\;\;\;\;\;\;\; m < \alpha < m+n \cr
i\left(\frac{m-n-1}{2}\right) \gamma \;\;\;\;\;\;\;\;\; \alpha=m+n \cr}$ \\ \hline
$osp(2n|2m)^{(1)}$ & $\cases{
i\frac{\alpha}{2}\gamma \;\;\;\;\;\;\;\;\;\;\;\;\;\;\;\;\;\;\;\; 1 \leq \alpha \leq m \cr
i\left(m -\frac{\alpha}{2}\right) \gamma \;\;\;\;\;\;\;\; m < \alpha \leq m+n-2 \cr
i\left(\frac{m-n+1}{2}\right) \gamma \;\;\;\;\;\;\;\;\; \alpha=\pm \cr}$ \\ \hline
\end{tabular}
\end{center}
\end{table}

\normalsize
\vspace{1.5cm}

{}

\end{document}